\documentclass[a4paper,fleqn,usenatbib]{mnras}
\usepackage{amsmath}
\usepackage{txfonts}
\usepackage{caption}
\usepackage[T1]{fontenc}
\usepackage{ae,aecompl}
\usepackage{hyperref}
\hypersetup{colorlinks, citecolor=green, filecolor=black, linkcolor=blue, urlcolor=blue }

\usepackage{graphicx}	% Including figure files	% Advanced maths commands
\usepackage{amssymb}	% Extra maths symbols
\usepackage{dblfloatfix}
\usepackage{breqn}

\title[Energy Dependence of BHB Flows]{Modelling the Energy Dependence of Black Hole Binary Flows}

\author[Mahmoud \& Done]{
Ra'ad D. Mahmoud$^{1}$\thanks{E-mail: ra'ad.d.mahmoud@durham.ac.uk} \&
Chris Done$^{1}$
\\
$^{1}$Department of Physics, University of Durham, South Road, Durham DH1 3LE\\
}

\date{Accepted XXX. Received YYY; in original form ZZZ}

\pubyear{2017}

\hypersetup{draft}
\begin{document}
\label{firstpage}
\pagerange{\pageref{firstpage}--\pageref{lastpage}}
\maketitle

% Abstract of the paper
\begin{abstract}
We build a full spectral-timing model for the low/hard state of black hole binaries assuming that the spectrum of the X-ray hot flow can be produced by two Comptonisation zones. Slow fluctuations generated at the largest radii/softest spectral region of the flow propagate down to modulate the faster fluctuations produced in the spectrally harder region close to the black hole. The observed spectrum and variability are produced by summing over all regions in the flow, including its emission reflected from the truncated disc. This produces energy-dependent Fourier lags qualitatively similar to those in the data. Given a viscous frequency prescription, the model predicts Fourier power spectral densities and lags for any energy bands. We apply this model to archival RXTE data from Cyg X-1, using the time-averaged energy spectrum together with an assumed emissivity to set the radial bounds of the soft and hard Comptonisation regions. We find that the power spectra cannot be described by any smooth model of generating fluctuations, instead requiring that there are specific radii in the flow where noise is preferentially produced. We also find fluctuation damping between spectrally distinct regions is required to prevent all the variability power generated at large radii being propagated into the inner regions. Even with these additions, we can either the power spectra at each energy, or the lags between energy bands, but not both. We conclude that either the spectra are more complex than two zone models, or that other processes are important in forming the variability.

\end{abstract}


%%%%%%%%%%%%%%%%%%%%%%%%%%%%%%%%%%%%%%%%%%%%%%%%%%

%%%%%%%%%%%%%%%%% BODY OF PAPER %%%%%%%%%%%%%%%%%%

\section{Introduction}
\label{Introduction}
Black hole binaries (BHBs) show variability on a wide range of timescales. Over days, months and years, mass accretion rate changes drive changes in the energy spectrum. The most dramatic example of this behaviour is the spectral transition from the Comptonisation-dominated (low/hard) spectra seen at low luminosities to the disc-dominated (high/soft) spectra at high luminosities (see e.g. \citealt{RM06}). This has a very natural interpretation from the two stable solutions to the accretion flow equations: one which is hot, optically thin and geometrically thick (advection dominated accretion flow, ADAF; \citealt{NY95}) which can only exist at low mass accretion rates, and one which is cool, optically thick and geometrically thin (Shakura-Sunyaev disc, SS; \citealt{SS73}). The observed switch in spectral properties can therefore be explained by a switch between these two solutions at the maximum ADAF luminosity (\citealt{EMN97}; \citealt{DGK07}; hereafter DGK07).

There is also more subtle spectral evolution within the low/hard state. This can be explained by combining these two solutions into a composite structure, where the outer SS disc truncates at some radius to be replaced by a hot flow interior to this (truncated disc/hot flow models). In this geometry, decreasing the truncation radius as the mass accretion rate increases results in more disc seed photons incident upon the flow. This leads to more efficient Compton cooling of the flow, and naturally produces the softer Comptonised spectra with increasing luminosity, as observed (DGK07).

However, the broadband spectra show more complexity than the contributions from a simple truncated disc, a single-temperature Comptonisation region, and its reflection from that disc. This complexity can be fit by assuming that the Comptonisation is not at a single-temperature, but instead is radially stratified. Simple inhomogeneous flow models consisting of a truncated disc and two Comptonisation components can broadly fit the 0.2-200~keV spectra seen in the low/hard states of Cyg X-1 (\citealt{G97}; \citealt{DS01}; \citealt{M08}; \citealt{Y13}; \citealt{B17}).

Alternatively, the additional X-ray component in low/hard state spectra can instead be modelled by a jet contribution (\citealt{MNW05}; \citealt{N11}), or using the completely different geometry of an untruncated disc with highly relativistic reflection from a point-source on the spin axis of the black hole (e.g. \citealt{Ry07}; \citealt{RFM10}; \citealt{F14}). Variations on the theme of the truncated disc/hot flow model where some of the hot flow electrons have a hybrid (thermal/non-thermal) electron distribution have also successfully fit the spectra (e.g. \citealt{PC98}; \citealt{I05}; \citealt{M08}; \citealt{PV09}; \citealt{N11}). However, the fast timing (0.01-100~s) properties can break some of these spectral degeneracies by giving additional information on the source geometry. In particular, the evolution of the power spectral density (PSD) of the fast timing variability in the low/hard state strongly supports the truncated disc/hot flow geometry (\citealt{ID11}, hereafter ID11). The PSD of the Compton-dominated X-ray emission shows band-limited noise between low ($f_b$) and high ($f_h$) frequency breaks, often accompanied by a strong low-frequency quasi-periodic oscillation (QPO) at $f_{qpo}$. Both $f_b$ and $f_{qpo}$ increase together as the spectrum softens towards the transition (\citealt{WvdK99}; \citealt{KWvdK08}; \citealt{RIvdK14}), indicative of the decreasing characteristic radius predicted by the truncated disc models.

This geometry can be incorporated into a full timing model by assuming that density fluctuations are generated at all radii in the hot flow by the turbulent magnetic dynamo (magneto-rotational instability: MRI, \citealt{BH98}). These fluctuations propagate inwards on the viscous timescale, so that slow fluctuations stirred up at large radii modulate the faster fluctuations generated at smaller radii (\citealt{L97}). This process can reproduce the observed double-broken power law shape of the low/hard state PSD, while Ar\'{e}valo \& Uttley (2006; hereafter AU06) also show that this behaviour is necessary and sufficient to produce the observed linear rms-flux relation. The correlated QPO can also be produced from the same geometry if the entire hot flow undergoes Lense-Thirring precession due to its misalignment with the black hole spin axis (\citealt{FM09}; \citealt{IDF09}; \citealt{L17}). These propagating fluctuation/Lense-Thirring precession models have quantitatively fit the data from XTE J1550-584 during its spectral transition, with the inner radius of the thin disc changing from $\sim 60-12R_g$ (ID11; \citealt{ID12a}, hereafter ID12a), while also correctly predicting the modulation of the iron line energy on the QPO period (\citealt{ID12b}; \citealt{I16}).

Thus, the overall properties of the power spectra already strongly favour the truncated disc/hot flow geometry for the low/hard state, but they do not break the degeneracies between the different models for the X-ray emission within this framework. However frequency-dependent time lags are also observed between high and low-energy X-ray bands (\citealt{MK89}; \citealt{N99}). These were first discovered in data above 2~keV, so they directly probe the structure of the hot flow rather than the disc emission.  The lags show that flux variations are seen first in the softer X-rays, and later in the hard X-rays (hard lags), after a lag time which depends on the fluctuation frequency. This frequency-dependence rules out a simple light travel time origin for the signal, such as the delay between successive Compton scattering orders, as this would produce a constant, very short hard lag (\citealt{MK89}; \citealt{N99}). The light travel time between the source and disc in the reflection dominated spectra is also ruled out as the source of the hard lag, as this process results in only a constant, very short soft lag.

Instead, the observed frequency-dependent hard lags can be qualitatively explained by the propagation of fluctuations through an inhomogeneous hot flow such as the one we have described, where the Compton spectrum is harder closer to the black hole (\citealt{KCG01}). This results in a coupled spectral-timing model, where slower fluctuations are produced at larger radii, so have softer spectra. These fluctuations propagate down to smaller radii, modulating the harder spectra from these regions. The lag time for this propagation encodes the viscous timescale between the radii. Faster fluctuations are produced at smaller radii, so they have a shorter distance to propagate before they modulate the hardest spectra from the innermost region, giving the frequency dependent lag time (AU06). By contrast, in a model where the soft X-rays are from the jet, fluctuations would propagate down through the accretion flow which produces hard Comptonisation, and only then propagate up into the jet which produces soft X-ray synchrotron. This would instead predict a soft lag, contrary to what is observed.

Hence the spectral lags impose additional constraints on the physical nature and geometry of the hot flow. Here we build a fully energy-dependent spectral-timing model of the simplest possible inhomogeneous hot flow interior to a truncated disc, where the flow is composed of only two Comptonisation regions of different temperature and optical depth. We quantitatively compare the predictions of this model to the best fast spectral-timing data currently available: the archival Rossi X-ray Timing Explorer (RXTE) observations of Cygnus X-1 in the low/hard state (see $\S$ \ref{Cyg X-1}), as used by \cite{N99}. The data span an energy range of 3-30~keV, so they are dominated by the emission from the hot flow, and exclude the truncated disc emission.

Model fitting to X-ray lags has only recently become possible, and our approach is complementary to the few papers produced so far on this. Rapisarda \textit{et al.}~(2016, hereafter R16) use lower energy data (0.5-10~keV) from Swift to model the power spectra in a soft and hard band, and the lags between them for the BHB MAXI J1659-152. This lower energy band means that they consider the intrinsic disc emission and its variability (\citealt{U11}) and how this propagates into the hot flow, which they assume is homogeneous. This model is adequate to describe that dataset as it does not extend above 10~keV.  However, the same disc and homogeneous hot flow model fails to fit the RXTE data from XTE J1550-564 (\citealt{R17}, hereafter R17), potentially because the higher energy range (2-30~keV) of these data mean that the cross-spectral properties are sensitive to the structure within the hot flow, and such structure is not incorporated into their model.

We describe the data we compare to in $\S$ \ref{Cyg X-1}, while $\S$ \ref{The Propagating Fluctuations Model} briefly details the single zone propagating fluctuation model. In $\S$ \ref{Extending} - \ref{SoftPower} we systematically build our procedure to predict frequency-dependent time lags, by applying different spectral components to different radial ranges in the propagating fluctuations model. Finally in $\S$ \ref{conclusions}, we discuss the successes and failures of our model prescription and directly tie these back to the nature and geometry of the X-ray emission region close to the black hole.

\section{Observations of Cygnus X-1 in the Hard State}
\label{Cyg X-1}
Cygnus X-1 is typically the brightest low/hard state source, and so gives the best data for studies using high time resolution. The archival data from RXTE remains the best publically available data for studying the Comptonisation lags, due to its high effective area in the 3-30 keV bandpass. Many of the RXTE observations were taken in a mode with limited spectral resolution below 10 keV. However, there are 6 datasets taken in the `Generic Binned' mode which has 15.6 millisecond time resolution with 64 energy bins across the entire RXTE PCA energy bandpass (standard channels 0-249; {\tt{B_16ms_64M_0_249}} configuration) giving reasonable spectral resolution in the 3-10 keV band, which allows the broad iron line to be resolved (\citealt{RGC99}; \citealt{GCR00}).

We use three of these observations taken consecutively during 1996, with simultaneous data from the Proportional Counter Array (PCA) and the High Energy X-Ray Timing Experiment (HEXTE; ObsIDs: 10238-01-08-00, 10238-01-07-000, 10238-01-07-00, hereafter observations 1-3). We choose these as they have very similar time averaged spectra, with hardness ratios between the 6-10 and 3-6 keV bands of $0.9151~\pm~0.0003$, $0.9149~\pm~0.0004$ and $0.9148~\pm~0.0003$ respectively. The remaining three observations in this mode are all somewhat softer, so we exclude them. All 5 Proportional Counter Units (PCUs) of the PCA were active during these epochs. Each observation is background-subtracted (using background on 16 s time binning), Poisson noise is removed, and dead-time corrections are applied according to the standard procedure of \cite{N99}.

Observations 1-3 also have statistically consistent power spectra at the $1\sigma$ level across the entire frequency range, so we co-add these observations to give 22.5ks of data for the timing analysis. However we use only Obs. 1 for spectral analysis, as the co-addition of spectral data with slightly different response matrices can lead to artefacts. 

Even amongst observations restricted to the hard state, a range of `sub-states' are seen in both the variability and the spectra (e.g. the hard-intermediate state; DGK07). We would therefore like to place our observations in the wider context of states seen from Cygnus X-1. Grinberg \textit{et al.} (2014; hereafter G14) fit all the Cyg X-1 data taken during the lifetime of RXTE with a phenomenological model of {\tt{tbabs*(gaussian + highecut*bknpower)}}, where the \linebreak{\tt{bknpower}} component approximates the Comptonised emission as a broken power law, parameterised by ``soft" and ``hard" photon indices, $\Gamma_1$ and $\Gamma_2$ respectively. Our data has a ``soft" photon index of $\Gamma_1 = 1.65 \pm 0.01$, which is the minimum $\Gamma_1$ found by G14, showing that this is one of the hardest states of Cyg X-1 observed by RXTE. This extreme hard state is confirmed by the high fractional root-mean-square variability (\citealt{MMB11}; \citealt{HVU12}) in the 2-15~keV band of $26.3 \pm 0.5\%$.

For our analysis we use lightcurves in three energy bands: Low (3.13-4.98~keV), Mid (9.94-20.09~keV) and High (20.09-34.61~keV). We extract these using {\sc{saextrct}}, ensemble averaging over 174 segments of 128~s length to derive power spectra and time lags which are far better constrained at high frequencies than previous model-comparison studies (R16; R17).

\section{The Propagating Fluctuations Model}
\label{The Propagating Fluctuations Model}
The magneto-rotational instability (MRI) threading the flow generates fluctuations in all quantities and on all timescales (\citealt{BH98}). The stochastic variations in mass accretion rate propagate down through the Comptonising region, modulating all the faster fluctuations produced further in. We simulate a Comptonisation region extending from the thin disc truncation radius, $r_o$, to the inner edge of the hot flow at $r_i$, where all size scales are in units of $R_g = GM/c^2$. The flow is split into annuli, characterized by radius $r_n$ and width $dr_n$, logarithmically spaced such that $dr_n/r_n$ is constant (ID11).

The largest amplitude fluctuations produced by any given radius have size $\sim h$, where $h$ is the thickness of the flow. For $r\sim h$, this sets the local viscous time, $t_{visc}(r)$, as the shortest timescale on which density fluctuations are generated at $r$; fluctuations on shorter timescales than this are damped by the response of the flow. This results in a break in the power spectrum of mass accretion rate fluctuations generated at $r$ of $f_{visc}(r)=1/t_{visc}(r)$ (\citealt{KCG01}). The largest radius in the flow generates the slowest fluctuations, so the low-frequency break in the observed PSD is $f_{visc}(r_o)$.

However, translating this to an outer radius requires a functional form for the viscous timescale. This form is not yet clear. General Relativistic Magneto-Hydrodynamical (GRMHD) simulations of the MRI currently predict that fluctuations can be generated on ten-times the Keplerian timescale, $\sim10t_{kep}(r)$ (\citealt{HR17}). However this predicts that the typical low-frequency break seen in hard-state power spectra at $\sim0.1$~Hz is produced by material at large distances, of order several hundred $R_g$. This is in tension with results from spectral fitting to the iron line profile, which generally point to $r_o \lesssim 50$ (\citealt{KDD14}; \citealt{B17}). This inconsistency is likely due to the limited physics currently incorporated into the GRMHD simulations. Typically these neglect radiative processes and the interface between the disc and hot flow (e.g. \citealt{L17}). Until better simulations are available, we instead use a parameterised prescription where $f_{visc}(r) = Br^{-m} f_{kep}(r)$ (ID11). For a thin SS disc, $m=0$ and $B=\alpha(h/r)^2$ (with $h/r<< 1$) while a self-similar ADAF adheres to the same scalings but with $h\sim r$. More complex flows have $m\ne0$, for example when including transonic effects in an ADAF (\citealt{NKH97}) or in full MRI simulations (\citealt{FM09}).

We follow ID11, who determine $B$ and $m$ from fitting to the well known relationship between $f_{qpo}$ and $f_b$ (\citealt{WvdK99}). This gives $B=0.03$ and $m=0.5$, assuming that the QPO is indeed from Lense-Thirring precession of the entire hot flow and that $f_b\approx f_{visc}(r_o)$. This is a simpler prescription than using a GRMHD surface density to derive $f_{visc}(r)$ as in ID12a and R16/17, and avoids the associated simulation uncertainties.

We assume that these stochastic mass accretion rate fluctuations are generated at each radius $r_n$ with random phase, but with a well defined power spectrum which is a zero-centered Lorentzian with a cut-off at $f_{visc}(r_n)$, 
\vspace*{-3 pt}
\begin{equation}
\label{eq:Lorentzian}
|\tilde{\dot{m}}(r_n, f)|^2 \propto \frac{1}{1+[f/f_{visc}(r_n)]^2} \left[\frac{sin(\pi f dt)}{\pi f dt}\right]^2,
\end{equation}
where a tilde denotes the Fourier transform. The sinusoidal term on the right hand side describes the suppression of variability due to the time binning. The normalisation of our Lorentzian is selected such that all $\dot{m}(r_n, t)$ have a mean of $\mu=1$ and fractional variability $\sigma/\mu = F_{var} / \sqrt{N_{dec}}$, where $N_{dec}$ and $F_{var}$ are the number of annuli and fractional variability generated per radial decade respectively.

Beginning at the outermost annulus, $r_1=r_o$, we generate mass accretion rate fluctuations in the time domain, $\dot{m}(r_n, t)$ using the algorithm of Timmer and K{\"o}nig (1996). For the outermost annulus we designate the accretion rate across the annulus as \linebreak$\dot{M}(r_1, t)=\dot{M}_{0} (1+\dot{m}(r_1, t))$ where $\dot{M}_{0}$ is the mean mass accretion rate. These fluctuations propagate in to the next annulus, traveling a distance $dr_{1}$, which takes a time $d\tau_1=dr_{1}/[r_1 f_{visc}(r_1)]$.

The response of the flow acts to smooth fluctuations on the lag timescale. We implement this via a moving average over the lag time across the light curve, such that the smoothed mass accretion rate is
\begin{equation}
\label{smoothing}
\dot{M}_{sm}(r_n, t)
= \frac{\sum\limits_{t_i = t - d\tau_n/2}^{t + d\tau_n/2} {\dot{M}(r_n, t_i)}}{d\tau_{n}/dt}.
\end{equation}

Taken together with time lags, the total propagated mass accretion rate function in the $n^{th}$ annulus is then
\vspace*{-3 pt}
\begin{equation}
\label{pureprop}
\dot{M}(r_n, t)
= \dot{M}_{sm}(r_{n-1}, t - d\tau_{n-1}) [1+\dot{m}(r_n, t)],
\end{equation}
until the $N^{th}$ annulus which is $r_i$. These mass accretion rate functions are the fundamental quantity in several previous studies \linebreak (e.g. AU06; ID11; ID12a) which accurately replicate the broken power law shape in BHB power spectra.

These works conventionally convert the mass accretion rate curves to light curves via $dL(r_n, t) = 0.5 \dot{M}(r_n, t)\epsilon(r_n)r_n dr_n c^2$ where $\epsilon(r_n)$ is the emissivity at annulus $r_n$ which can be parameterised in a number of ways depending on the assumptions made regarding energy dissipation. Instead, a key extension of our work is that the total energy dissipation is set by the gravitational energy release, with the photon energy dependence set by the different spectra generated at different radii.
\begin{figure}
	\includegraphics[width=\columnwidth]{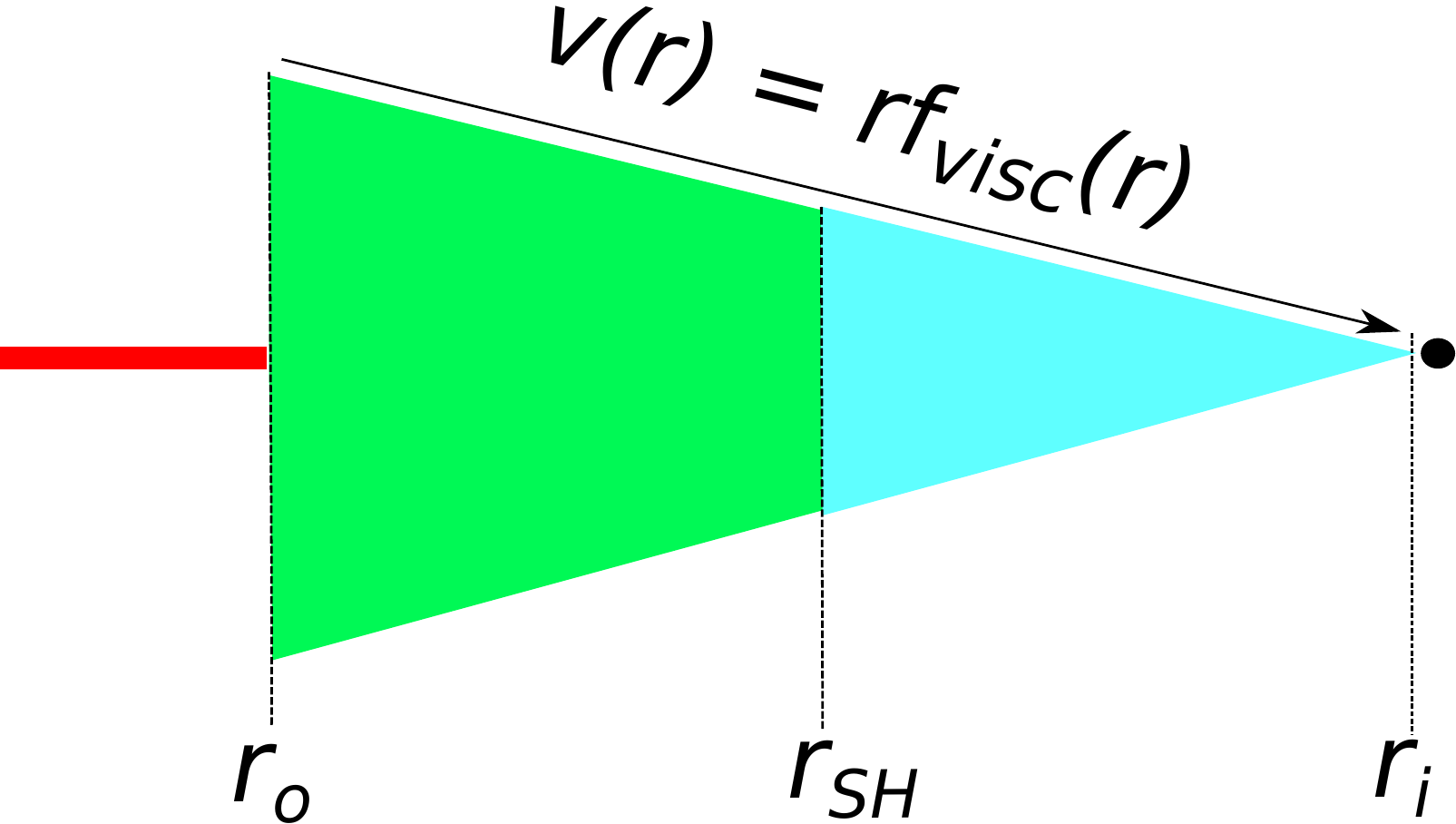}%
	\caption{The assumed geometry. The green region emits the soft spectral shape, $S(E)$, and the cyan region emits the hard spectral shape, $H(E)$. The direct contribution from the thermal disc (red) is neglected. Fluctuations are generated throughout the flow on the viscous timescale, $t_{visc}(r)$, and propagate down toward the compact object at a local velocity, $v(r)=rf_{visc}(r)$.}
	\label{fig:2compgeo}
\end{figure}

\section{Incorporating Energy Dependence}
\label{Extending}
\subsection{Spectral decomposition}
\label{SpecDec}

The standard propagating fluctuations model assumes a constant spectral energy distribution (SED) across the entire hot flow. It is only the normalisation of this SED which varies in time according to the variability of the flow at each radius, while the shape is assumed to be invariant. However physically there is more energy from gravitational heating of the flow at smaller radii, and fewer seed photons cooling it, so we expect the inner regions to have higher temperatures and hence harder spectra (\citealt{PV14}). Since the viscous frequency is also a function of radius, this couples the spectral and timing properties so that the cross-spectral statistics can be derived. This allows us to jointly compare the PSDs and time lags as a function of energy band as described by the propagating fluctuations model, \textit{simultaneously} with the time-averaged SED.

\begin{figure}
	\includegraphics[width=\columnwidth]{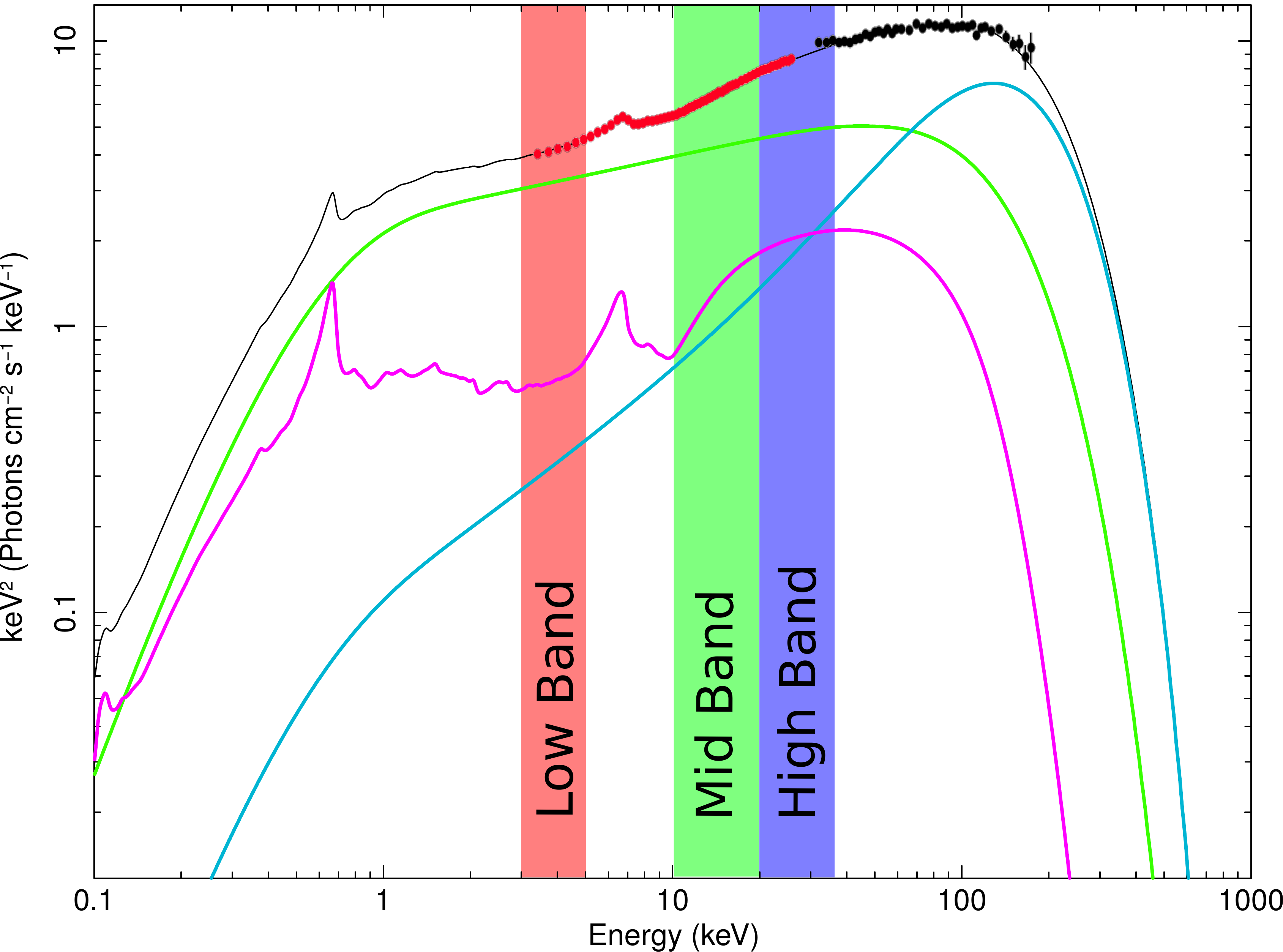}%
	\caption{The decomposition of Observation 1 (ObsID: 10238-01-08-00) used to augment the standard propagating fluctuations model. Shown are the total energy spectrum (black), the hard Compton component ($H(E)$, cyan), the soft Compton component ($S(E)$, green), and the reflection component ($R(E)$, magenta). Filled circles show the PCA (red) and HEXTE (black) data. The red, green and blue bands denote the Low (3.13-4.98~keV), Mid (9.94-20.09~keV) and High (20.09-34.61~keV) energy ranges respectively.}
	\label{fig:2compspec}
\end{figure}

The simplest multicomponent flow is described by two main Comptonisation regions: one softer component close to the disc, and one harder close to the black hole (see Fig.~\ref{fig:2compgeo}). We therefore fit the time averaged SED, with 0.5\% systematic errors, in {\sc{xspec}} (version 12.9.1; \citealt{ABH96}) with two Comptonisation components described
by {\tt{tbabs*(nthcomp+nthcomp)}} (\citealt{ZJM96}), and the combined reflection of these, {\tt{tbabs*(kdblur*xilconv*twocomp)}}. Here {\tt{twocomp}} is a local model which adds the Comptonisation components together, so that reflection is explicitly calculated from the composite spectrum. Such a decomposition is motivated both by model simplicity, and by similar successful fits to Cyg X-1 spectra (\citealt{G97}; \citealt{DS01}; \citealt{M08}; \citealt{B17}). We also follow \citet{M08} and assume that both Compton components have the same electron temperature. The data and best fit model are shown in Fig.~\ref{fig:2compspec}, with full parameters detailed in Table~\ref{tab:specparams}. The softer and harder Comptonisation components, $S(E)$ (green) and $H(E)$ (cyan), originate from the outer and inner regions of the flow respectively. Also included is the total reflection from the disc, $R(E)$, but we do not include the intrinsic or reprocessed disc emission as the energy of this is too low to make a significant contribution to the RXTE data above 3~keV.

\begin{table}
	\centering
	\begin{tabular}{lcccccccccccccr}
		\hline
		Component & Parameter & Value \\
		\hline
		\vspace{+3pt}
		nthComp & $\Gamma$ & $1.795^{+0.001}_{-0.005}$  & \\
		\vspace{+3pt}
		& $kT_e$ (keV) & $44^{+1}_{-2}$ &\\
		\vspace{+3pt}
		& norm & $2.2^{+0.7}_{-0.2}$  &\\ 
		\vspace{+3pt}
		nthComp & $\Gamma$ & $1.25^{+0.02}_{-0.01}$& \\
		\vspace{+3pt}
		& $kT_e$ (keV) & $44^{+1}_{-2}$ &\\
		\vspace{+3pt}
		& norm & $0.07^{+0.03}_{0.01}$ &\\
		\vspace{+3pt}
		xilconv & relative refl norm & $-0.254 \pm0.003$ & \\
		\vspace{+3pt}
		& log($x_i$) & $3.001^{+0.007}_{-0.005}$ & \\
		\hline
	\end{tabular}
	\caption[caption]{Fit parameters for the spectral model shown in Fig.~\ref{fig:2compspec}, with the reflected emission from the sum of the two Comptonisation components: {\tt{tbabs*(nthcomp+nthcomp+kdblur*xilconv*twocomp)}}. The electron temperatures ($kT_e$) of both Comptonisation components are tied. The fixed parameters in our fits are the galactic absorption column density ($0.6 \times 10^{22}$ cm$^{-2}$), the seed photon temperature (0.2 keV), the {\tt{kdblur}} index (3.0), the inclination angle of Cyg X-1 (27$^o$), the inner disc radius (10 $R_g$), and the {\tt{xilconv}} iron abundance (1.0).}
	\label{tab:specparams}
\end{table}

\subsection{Spectral-timing model}
\label{ID11MwED}

\begin{figure}
	\includegraphics[width=\columnwidth]{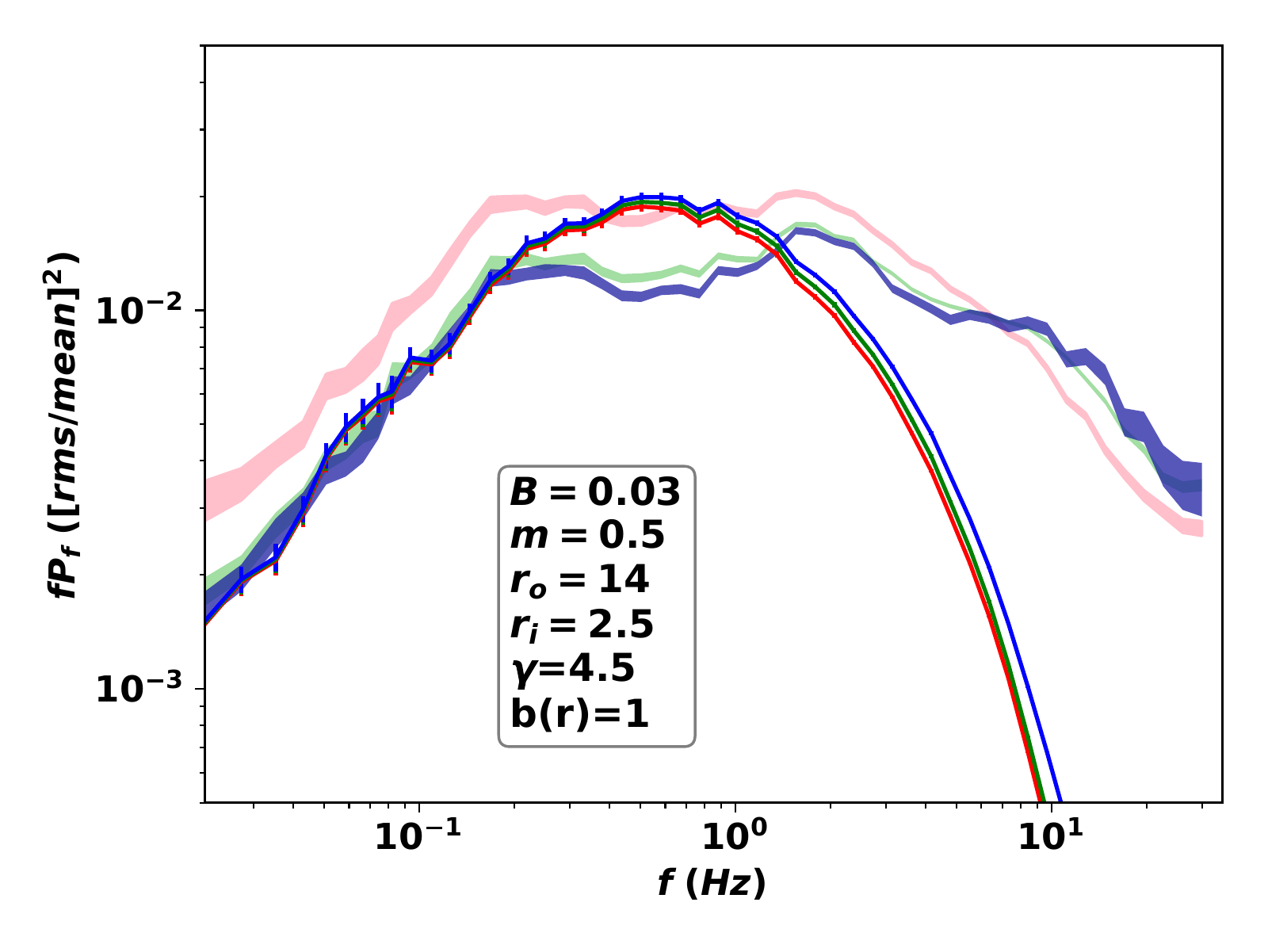}
	\caption{PSDs for the data, and for the energy-dependent ID11 model with $\gamma = 4.5$, $b(r) = 1$. The shaded regions are the 1$\sigma$ error regions of the Low (pink), Mid (green) and High (blue) energy bands from the data. The solid lines show the Low (red), Mid (green) and High (blue) energy model outputs.}
	\label{fig:ID11PSD}
\end{figure}

In all simulations we assume that Cyg X-1 has a black hole of mass, $M_{BH} = 15 M_{\odot}$, and a dimensionless spin parameter of $a^* \sim$ 0.85 (\citealt{K17}). The inner radius is set to the approximate ISCO size implied from the spin of Cyg X-1, so that $r_i = 2.5$.

The time-averaged spectrum, $\bar{F}(E,r_n)$, emitted from each radius is given by the expression
\vspace*{-3 pt}
\begin{equation}
\label{F_r}
\bar{F}(E, r_n)=
  \begin{cases}
    S(E)[1+\frac{R(E)}{S(E)+H(E)}] & \text{if}\ r_n > r_{SH}, \\
    H(E)[1+\frac{R(E)}{S(E)+H(E)}] & \text{if}\ r_n < r_{SH}.
  \end{cases}
\end{equation}
$r_{SH}$ here is the transition radius between the soft and hard Comptonisation regions. This is analytically derived from an assumed emissivity, $\epsilon(r)\propto r^{-\gamma} b(r)$, where $b(r)$ is an inner boundary condition, such that the luminosity ratio between the two components matches that observed, such that
\vspace*{-3 pt}
\begin{equation}
\label{r_SH}
\frac {\int_E S(E) dE}{\int_E H(E) dE} = \frac {\int_{r_o}^{r_{SH}} \epsilon(r)
  2\pi r dr }{\int_{r_{SH}}^{r_i} \epsilon(r)  2\pi r dr }.
\end{equation}

\begin{figure*}
	\includegraphics[width=\textwidth]{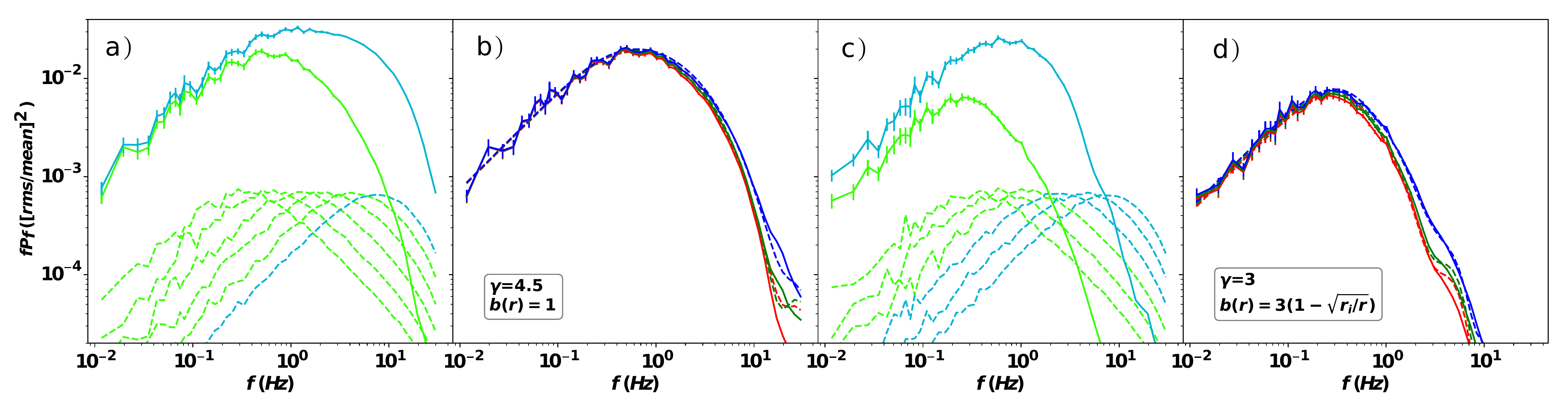}%
	\caption{(a): Dashed lines are rms-normalised generator PSDs from separate annuli, log-spaced within the flow. Only 5 are shown for graphical clarity. Green colour denotes those from outer region ($r_n>r_{SH}$). Cyan colour denotes those in inner region ($r_n < r_{SH}$). Solid lines are the rms-normalised PSDs of the outer (green) and inner (cyan) regions. Here we have $\gamma = 4.5$, $b(r)=1$. (b): Simulation and analytic prediction for $\gamma = 4.5$, $b(r)=1$ emissivity, showing Low (red), Mid (green) and High (blue) bands. Solid lines denote simulation output. Dashed lines denote analytic prediction. (c): As in (a), but with $\gamma = 3$, $b(r) = 3 (1-\sqrt{r_i/r})$. (d): As in (b), but with $\gamma = 3$, $b(r) = 3 (1-\sqrt{r_i/r})$.}
	\label{fig:semianalyticdemo}
\end{figure*}

The light curves produced by the standard propagating fluctuations model at each radius are then made energy-dependent and renormalised such that their time-average is the flux for that energy bin and radius, yielding
\vspace*{-1pt}
\begin{equation}
dF(E, r_n, t) = \bar{F}(E, r_n)\dot{M}(r_n,t)\frac{\epsilon(r_n) r_n dr_n}{\sum\limits_{region} {\epsilon(r_n) r_n dr_n}} dE. 
\label{eq:dL}
\end{equation}
The summation limits implied by `\textit{region}' are $\{r_o$~to~$r_{SH}\}$ for $r_n >r_{SH}$ and $\{r_{SH}$~to~$r_i\}$ for $r_n<r_{SH}$. This normalisation guarantees that if Eq. \ref{eq:dL} is time averaged and summed over all radii, the observed energy spectrum is reproduced.

We match our spectra to the data as closely as possible by converting these fluxes to count rates using the detector effective area $A_{eff}(E)$ and galactic absorption $N_H(E)$. The count rate is then expressed
\begin{equation}
\label{eq:CountRates}
dC(E, r_n, t) = dF(E, r_n, t)A_{eff}(E)e^{-N_H(E)\sigma_T},
\end{equation}
where $\sigma_T$ is the Thompson cross-section. 

In practice, Eq. \ref{eq:CountRates} describes a three-dimensional matrix, which can be operated on in different ways to obtain a variety of statistics. For instance, the total count rate in each energy band can be obtained by summing the matrix in Eq. \ref{eq:CountRates} over all radii, and over the energy band of interest, yielding
\vspace*{-3pt}
\begin{equation}
\label{eq:Cband}
C_{band}(t) = \sum_{E = E_{band}^{min}}^{E_{band}^{max}}\sum_{r_n = r_{i}}^{r_{o}} {dC(E, r_n, t)}.
\end{equation}
We use this quantity to produce the model power spectral and cross-spectral statistics, which we then fit to their analogues from the data.

We first use the viscous model of ID11, i.e. a frequency prescription with $B=0.03$ and $m=0.5$ as discussed in $\S$ \ref{The Propagating Fluctuations Model}. Tying the viscous frequency at the outer radius to the low-frequency break in the data so that \linebreak $f_{visc}(r_o) = 0.3$~Hz $\approx f_{b}$, we obtain an outer radius of $r_o = 14$, which is consistent with the range of disc truncation radii found from spectral fitting of 13-20 $R_g$ (\citealt{B17}). The fiducial model of ID11 had an emissivity described by $\gamma = 4.5$ and a stressed inner boundary condition, $b(r)=1$. Coupling this with our decomposition of the time averaged spectrum through Eq. \ref{r_SH} gives $r_{SH} = 3.1$.

We calculate the light curves on a time binning of $dt = 15.6$ ms (matched to the timing mode resolution of RXTE) and simulate $T=128$ s for each realisation, ensemble averaging over $M=64$ realisations. All simulations use $N_r=50$ radial bins, and we require $F_{var} = 0.45$ in order to match the slope and amplitude of the low-frequency break. A summary of all parameter values used in the simulations in this paper can be found in Table \ref{tab:params}.

Fig.~3 shows the model PSD from this simulation, where it is clear that this a poor match to the data. Overall, all energy bands show far too little high-frequency power. The model also predicts that the PSDs of all energy bands are similar, while the data shows that the Low band dominates at all frequencies below $8$~Hz. We analytically explore the factors determining the shapes of these PSDs below.

\section{Analytic power spectral models}

The pioneering work of Ingram \& van der Klis (2013, hereafter IK13) show how the PSD can be analytically calculated by considering how propagated PSDs are constructed in Fourier-space, and how they are weighted by the emissivity in calculating the final count spectra. We will now adapt their procedure to reflect our simulations, including the light curve weightings according to the energy spectrum. 

In the following we denote the PSDs generated in annulus $r_n$, as
$P_{gen}(r_n)$ while those which are propagated from all outer annuli
down to $r_n$ are denoted $P_{prop}(r_n)$. Propagation causes the noise in $r_2$ (closer to the black hole) to be
modulated by the noise in $r_1$, lagged by the viscous
timescale. Since this lag time is small compared to the generator
timescale in $r_1$, then it is almost perfectly coherent between $r_1$
and $r_2$ so that the power is additive, and $P_{prop}(r_2)\approx
P_{gen}(r_1)+P_{gen}(r_2)$. Generalising this, the propagated PSDs are
described by
\begin{equation}
\label{eq:PSDgen2prop}
P_{prop}(r_n, f) = \sum_{m=1}^{n}P_{gen}(r_m,f),
\end{equation}
where the assumed self-similar nature of the fluctuations means that all the individual $P_{gen}(r_n, f)$ have the same amplitude. Fig.~\ref{fig:semianalyticdemo}a (dashed lines) shows the generator PSDs of the individual annuli, with the soft region in green and the hard in cyan. The solid green and cyan lines show the propagated PSDs of the total soft and hard regions respectively. This shows the clear difference in high-frequency extent of the PSDs of the two regions, with the hard region producing substantial additional power above $1$~Hz.

Our mass accretion rates are converted to counts in a given band using the emissivity prescription and SED decomposition described in $\S$ \ref{SpecDec}. This effectively weights the propagated mass accretion rate from each annulus by a factor, $w_n^{\,band}$, given by 
\vspace*{-3pt}
\begin{equation}
w_{n}^{\,band}(r_n) = \frac{\epsilon(r_n) r_n dr_n}{\sum\limits_{region} {\epsilon(r_n) r_n dr_n}}\sum_{E = E_{band}^{min}}^{E_{band}^{max}} \bar{F}(E, r_n)A_{eff}(E)e^{-N_H(E)\sigma_T} dE.
\end{equation}
The count spectrum for that band can then be written
\begin{equation}
C_{band}(t) = \sum_{n = 1} ^{N} w_n^{\,band} \dot{M}(r_n,t).
\end{equation}
Since the mean count rate of $\dot{M}(r_n, t)$ is normalised to $\dot{M}_0$, the mean count rate in a given energy band is then
\begin{equation}
\mu_C = \sum_{n = 1}^{N} \dot{M}_0 w_n^{band}.
\end{equation}

\begin{figure*}
	\includegraphics[width=\textwidth]{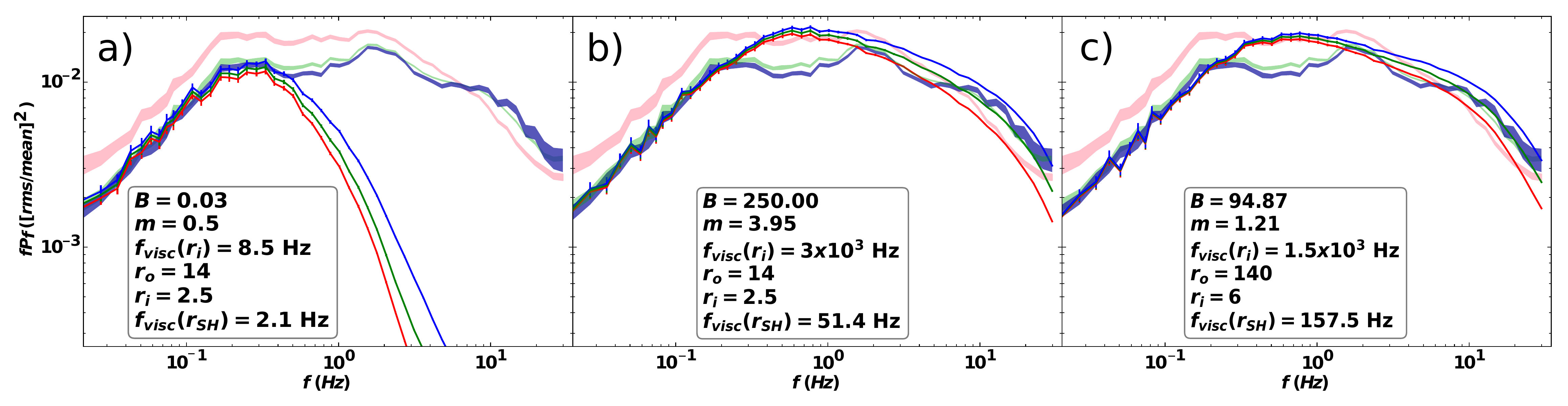}%
	\caption{Model and data PSDs. Colours as in Fig.~\ref{fig:ID11PSD}. Left (a): Fiducial model. As in Fig.~\ref{fig:ID11PSD} only now the emissivity prescription has been modified to $\gamma = 3$, $b(r) = 3(1-\sqrt{r_i/r})$. Middle (b): Viscous frequency parameters altered so that $f_{visc}(r_i)$ is now $3\times10^3$~Hz, allowing a match to the high-frequency power. Right (c): Parameters have been set such that the size scale is drastically different to (b) but the same PSD shape is found, illustrating the degeneracy between the frequency prescription and the radial range of the flow.}
	\label{fig:pureprop_PSDs}
\end{figure*}

We now drop the superscript on $w_n^{band}$ for notational convenience, and take the rms-normalised PSD:
\vspace*{-4pt}
\begin{equation}
\begin{aligned}
P_{band}(f) &= \frac{2dt^2}{\mu_{C}^2T}|\tilde{C}_{band}(f)|^2\\
&= \frac{2dt^2}{\mu_{C}^2T}\sum_{l,\,n=1}^{N} w_n w_l \tilde{\dot{M}}(r_l,f)^*\tilde{\dot{M}}(r_n,f).
\end{aligned}
\end{equation}

Including decoherence due to the propagation lag results in the full PSD expression of
\vspace*{-3pt}
\begin{equation}
\label{eq:PSDprop2band}
\begin{aligned}
P_{band}(f) = &\frac{1}{\mu_{C}^2}\sum_{n=1}^{N}\left[w_n^2 P_{prop}(r_n,f)\right.\vphantom{...}\\ 
&+ 2\sum_{l=1}^{n-1}\left.\vphantom{..} w_l w_n \text{cos}(2 \pi \Delta\tau_{ln} f)P_{prop}(r_l, f)\right], 
\end{aligned}
\end{equation}
where the weights now have a spectral dependence in our case. The second term in Eq.~\ref{eq:PSDprop2band} arises since the propagated noise at $r_n$ interferes with the propagated power spectra found at all outer radii. If this noise were not lagged between radii, this interference would be purely constructive and the cosine term would reduce to unity for all frequencies. However the time lag causes a component of the PSDs to interfere destructively, and this suppression is expressed by the lag-dependent cosine factor. Here, $\Delta\tau_{ln}$ is the total time lag between two annuli so that
\begin{equation}
\Delta\tau_{ln} = \sum_{m = l}^{n-1} d\tau_m = \sum_{m = l}^{n-1}\frac{dr_m}{r_m} t_{visc}(r_m)  = dlog(r) \sum_{m = l}^{n-1} t_{visc}(r_m),
\end{equation}
where the second equality comes from the fact that the radial bin size is logarithmic, and $dlog(r)$ is therefore a constant.

Eq.~\ref{eq:PSDprop2band} shows that each band-specific PSD is a weighted combination of the propagated PSDs in the outer and inner regions, where the weighting factor depends on the fraction of hard and soft spectral components in that band. Fig.~\ref{fig:2compspec} shows that the hard spectrum contributes more to the higher energy bands but is always a fairly small fraction of the total emission. Hence in Fig.~\ref{fig:semianalyticdemo}b (dashed lines), the analytic PSD of each band is dominated by the soft region. This results in band-dependent PSDs which are highly similar, with very little high-frequency power. We also show the full simulation output as the solid lines for comparison. Smoothing has been included in the simulations but is neglected in the analytic form, but the good agreement below $10$~Hz indicates that this effect is negligible.

It is clear that this combination of spectral decomposition, emissivity and viscosity will be unable to produce PSDs in each band which are close to the observations. Since the SED decomposition is fit prior to the spectral-timing model, we now explore how a better match to the power spectra can be achieved by varying the emissivity and viscous frequency prescription. We will start by reproducing the low-frequency Lorentzian hump at 0.2 Hz.

\begin{figure*}
	\includegraphics[width=\textwidth]{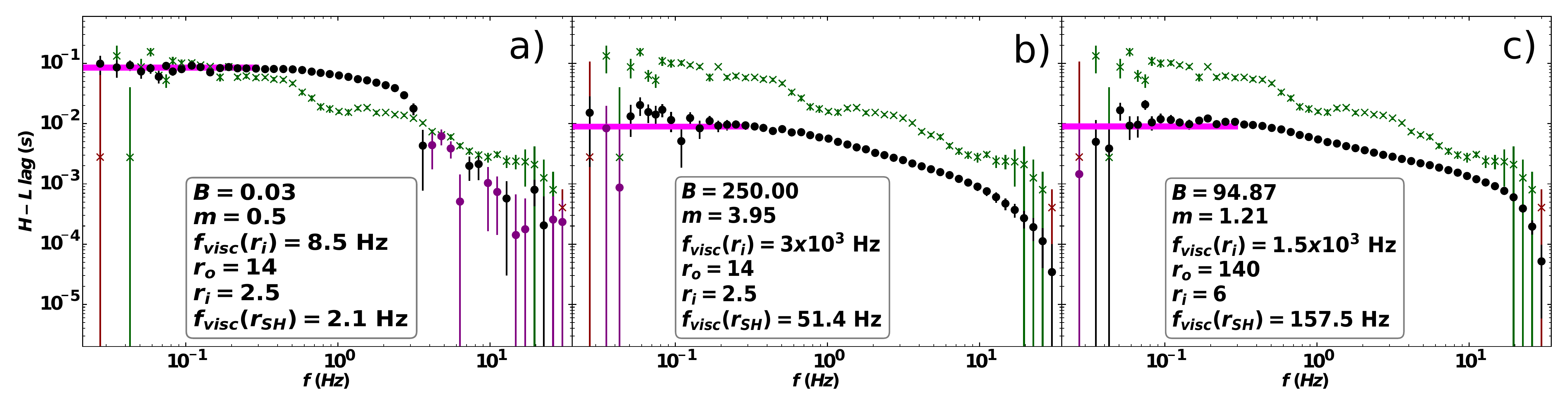}%
	\caption{High-Low band time lags for the data (crosses) and models (circles). Green (red) crosses indicate the High band lagging the Low band (or vice versa). Black (purple) circles indicate the High band lagging the Low band (or vice versa). Solid magenta lines indicate the analytic prediction for the low-frequency lag detailed in the text. Left (a): Fiducial model with $\gamma=3$, $b(r)=3(1-\sqrt{r_i/r})$. Corresponding PSDs in Fig.~\ref{fig:pureprop_PSDs}a. Middle (b): Viscous frequency parameters have been altered so that $f_{visc}(r_i)$ is now $3\times10^3$~Hz, allowing a match to the high-frequency power. Corresponding PSDs in Fig.~\ref{fig:pureprop_PSDs}b. Right (c): Viscous parameters set such that the size scale is drastically different to (b) but the same PSD shape is found, illustrating the degeneracy between the frequency prescription and the radial range of the flow. Corresponding PSDs in Fig.~\ref{fig:pureprop_PSDs}c.}
	\label{fig:pureprop_LAGS}
\end{figure*}

\section{Varying the emissivity and viscous frequency prescription}

\subsection{A physically motivated emissivity}
\label{Physically-Motivated}
Gravity gives an expected emissivity with $\gamma = 3$, while the innermost stable circular orbit motivates a stress-free (SF) inner boundary condition which we approximate as $b(r)=3(1-\sqrt{r_i/r})$. Together with the standard $B=0.03$, $m=0.5$ viscosity prescription, we will hereafter refer to this parameter set as the fiducial model. In Figs.~\ref{fig:semianalyticdemo}c and \ref{fig:semianalyticdemo}d we show the effect on the PSDs of applying this new emissivity, with a new $r_{SH} = 5.4$ set by the integrated SED components via Eq.~\ref{r_SH}. It is clear when comparing Figs.~\ref{fig:semianalyticdemo}a and c, that power at high frequencies has been suppressed in both the soft (green) and hard (cyan) bands. This is because the new emissivity weights the emission to larger radii, so the high-frequency contribution to the variability from the smallest radii is decreased. We also see that the simulated soft region power is now even lower than that from the hard region, even at frequencies below $0.3$~Hz.

In Fig.~\ref{fig:pureprop_PSDs}a, we fit the model with this new emissivity to the data, using $F_{var} = 0.59$ to match to the low-frequency break amplitude, but keeping all other parameters the same. This matches very well to the low-frequency PSD hump in the Mid and High bands, although it does not match the significantly higher amplitude of the Low band since this remains a total propagation model. However, the rest of the power spectrum is completely unmatched as the new, less centrally peaked emissivity means that more of the emission arises from larger radii, so the PSD is weighted more to lower frequencies. In effect, the emissivity defines an envelope which suppresses all power above 0.5-1~Hz (see the Appendix of AU06 for an energy-independent treatment).

\subsection{Different viscous frequencies, same radial range}
\label{ExtendedFreq}
The viscous parameterisation of ID11 assumed so far gives a maximum possible peak frequency of $f_{visc}(r_i) = 8.5$~Hz, although the finite width of the Lorentzians means that there is some power of even higher frequency generated near $r_i$. However, this high-frequency variability is suppressed, as the emissivity profile prevents these radii from producing a significant proportion of the total luminosity.

Increasing the maximum viscous frequency associated with the flow from $r_i$ would instead allow the PSD to extend to higher frequencies, while maintaining a gravitational emissivity. By diverging from $B=0.03$, $m=0.5$, we will break the $f_{QPO}$-$f_b$ relation, but we nevertheless explore this in order to better understand the effects of varying the viscous frequency prescription.

We first maintain the size scale of the region \linebreak($r_o=14$,~$r_i=2.5$) and emissivity ($\gamma=3$ with the SF boundary condition) so $r_{SH}$ stays constant at $5.4$, but we now fit $B$ and $m$ such that the PSD amplitudes at $f>8$~Hz are approximated. This yields $B=250.00$ and $m=3.95$. This keeps $f_{visc}(r_o)$ tied to $f_b$, but now gives $f_{visc}(r_i)=3\times10^3$~Hz. We see in Fig.~\ref{fig:pureprop_PSDs}b that, although it cannot match the peak structure seen in the data, this viscosity prescription can produce the observed high-frequency power. However, it has no physical motivation.

\subsection{ADAF viscous frequencies, large radial range}

The transonic ADAF models do make physical predictions about $f_{visc}(r)$, predicting $B=94.87$ and $m=1.21$ for $\alpha=0.1$ (\citealt{NKH97}). The very high ion temperature of the ADAF means that the sound speed and hence the radial velocity is high, so a much larger radial scale is required to produce the low-frequency break observed. We find a best fit of $r_o=140$ and $r_i=6$, which gives $r_{SH}=16$. The PSDs produced by this very different parameter set (Fig.~\ref{fig:pureprop_PSDs}c) are indeed equivalent in all essential features to those of the standard size scale assumed in Fig.~\ref{fig:pureprop_PSDs}b, due to the similar viscous frequency ranges spanned by the models. 

This is a key degeneracy. Without any external information to set the viscous frequency prescription (such as assuming that the QPO is set by Lense-Thirring precession) then the size scale of the region cannot be determined from the PSD. The data do show time lags between bands, however, so we now explore whether those time lags can break this degeneracy.

\section{Time Lags}

So far we have only investigated which elements of the observed PSDs can be replicated by this energy dependent model. However cross-spectral statistics including time lags can also be extracted from our simulations. These give additional information to that contained in the power spectrum; a good match to the PSDs does not necessarily imply a good fit to the cross-spectral lag (and vice versa), so any complete energy-dependent model must match both the power spectra and energy spectrum simultaneously with the time lags.

Figs.~\ref{fig:pureprop_LAGS}a-c show the time lags for each of the three power spectra shown in Figs.~\ref{fig:pureprop_PSDs}a-c, all of which used the physically motivated emissivity with $\gamma=3$ and the SF inner boundary condition. It is striking that the fiducial prescription (a:~$B=0.03$,~$m=0.5$,~$r_o=14$,~$r_i=2.5$), which has an excellent match to the low-frequency Mid and High-band power spectra (Fig.~\ref{fig:pureprop_PSDs}a), also has an excellent match to the low-frequency lags (Fig.~\ref{fig:pureprop_LAGS}a). This frequency prescription came from fitting the $f_{QPO}-f_{b}$ relation in ID11, so the good match to the low-frequency lag amplitude therefore provides additional support for the assumed Lense-Thirring origin of the QPO. However, this model completely fails to match the lags above 2~Hz, because frequencies with $f>f_{visc}(r_{SH})\approx2$~Hz are produced only in the hard region. For frequencies above 2~Hz, the variability contribution in both bands therefore comes entirely from the hard region, so there are no spectral lags even though the fluctuations themselves are lagged.

Conversely, the viscosity prescription which gives higher frequencies over the same size scale can mostly match the PSD (Fig.~\ref{fig:pureprop_PSDs}b) but underpredicts the lags at \textit{all} frequencies (Fig.~\ref{fig:pureprop_LAGS}b). This underprediction in the lag occurs because the viscous speed in the flow goes as $v(r)=rf_{visc}(r)$. Compared to the fiducial prescription, $f_{visc}(r)$ is now higher (so $v(r)$ is faster) for all $r$, resulting in shorter lags. This prescription does produce significant lags up to a higher frequency however. This is because we now have $f_{visc}(r_{SH})=50$~Hz, so fluctuations slower than this are found in both the soft and hard regions, giving measurable lags at these frequencies.

The larger size scale ADAF model (Fig.~\ref{fig:pureprop_LAGS}c) produces very similar lags to those of Fig.~\ref{fig:pureprop_LAGS}b, highlighting the fact that degeneracies on size scale can remain even when incorporating time lags. We illustrate here why this occurs using a physically intuitive derivation of the maximum lag between the High and Low bands, but in Appendix~\ref{AppendixCrossSpec} we extend this to all frequencies using the formalism of IK13.

The radial velocity $v(r)$ is not constant, so the raw time lag, $\tau_0\neq rR_g/f_{visc}(r_o)$. The lowest frequency component, $f_o=f_{visc}(r_o)$, propagates down through the entire flow. Light curves are calculated by weight-summing over the flow, and all fluctuations therefore \textit{appear} to initiate at some radius $\left<r_S\right>$ as seen in the Low band, and arrive some time later at some radius $\left<r_H\right>$ as seen in the High band. $\left<r_S\right>$ and $\left<r_H\right>$ are the emissivity-weighted averages of all radii in the soft and hard regions respectively, so that 
\vspace*{-2pt}
\begin{equation}
\label{raves}
\left<r_S\right>=\frac{\int_{r_{SH}}^{r_o}r^2 \epsilon(r)dr}{\int_{r_{SH}}^{r_o}r \epsilon(r)dr},\,\,\,\ \left<r_H\right>=\frac{\int_{r_i}^{r_{SH}}r^2 \epsilon(r)dr}{\int_{r_i}^{r_{SH}}r \epsilon(r)dr}.
\end{equation}

The maximum raw lag is then the propagation time between these radii
\begin{equation}
\label{eq:lagmax}
\begin{aligned}
\tau_0 &= \int_{\left<r_H\right>}^{\left<r_S\right>} \frac{dr}{rf_{visc}(r)} \\
&= \frac{2\pi R_g}{Bc}\left[\frac{\left<r_S\right>^{m+3/2}}{m+3/2} - \frac{ \left<r_H\right>^{m+3/2}}{m+3/2} + \frac{\left<r_S\right>^m}{m} -\frac{\left<r_H\right>^m}{m}\right].
\end{aligned}
\end{equation}
This lag is then diluted by the soft SED contribution in the High band and the hard SED contribution in the Low band (\citealt{U14}), so we obtain
\begin{equation}
\label{taumax}
tan[2\pi f_o \tau_{dil}]=\frac{\text{sin}(2\pi f_o \tau_0)(1-X_L X_H)}{X_H + X_L + \text{cos}(2\pi f_o \tau_0)[1+X_L X_H]},
\end{equation}
where $X_H$ is the ratio of integrated soft flux to integrated hard flux in the High band and $X_L$ is the ratio of integrated hard flux to integrated soft flux in the Low band. The predicted $\tau_{dil}$ values for the maximum lags in Figs.~\ref{fig:pureprop_LAGS}a, b and c (indicated by the solid magenta lines) are 0.09, 0.009 and 0.009 s respectively, which are generally consistent with the simulation results. Higher frequencies demand a full cross-spectral treatment since they are more prone to interference (see Appendix~\ref{AppendixCrossSpec} and IK13), however this simple result confirms that two very different size scales/viscosity prescriptions can predict indistinguishable lags.

Fundamentally, the larger size scale and higher velocity of the ADAF prescription (Fig.~\ref{fig:pureprop_LAGS}c) is degenerate with the smaller size scale and lower velocity of the Fig.~\ref{fig:pureprop_LAGS}b prescription, as both are tuned to match the breaks in the PSD. This is a direct consequence of assuming that the fluctuations are generated \textit{and} propagated on the $t_{visc}(r)$ timescale. However, there are some models which do not require that the propagation and generation timescales are the same. We explore these below.

\subsection{Decoupling the timescales of propagation and generation}
\label{Timescales}
\begin{figure}
	\includegraphics[width=\columnwidth]{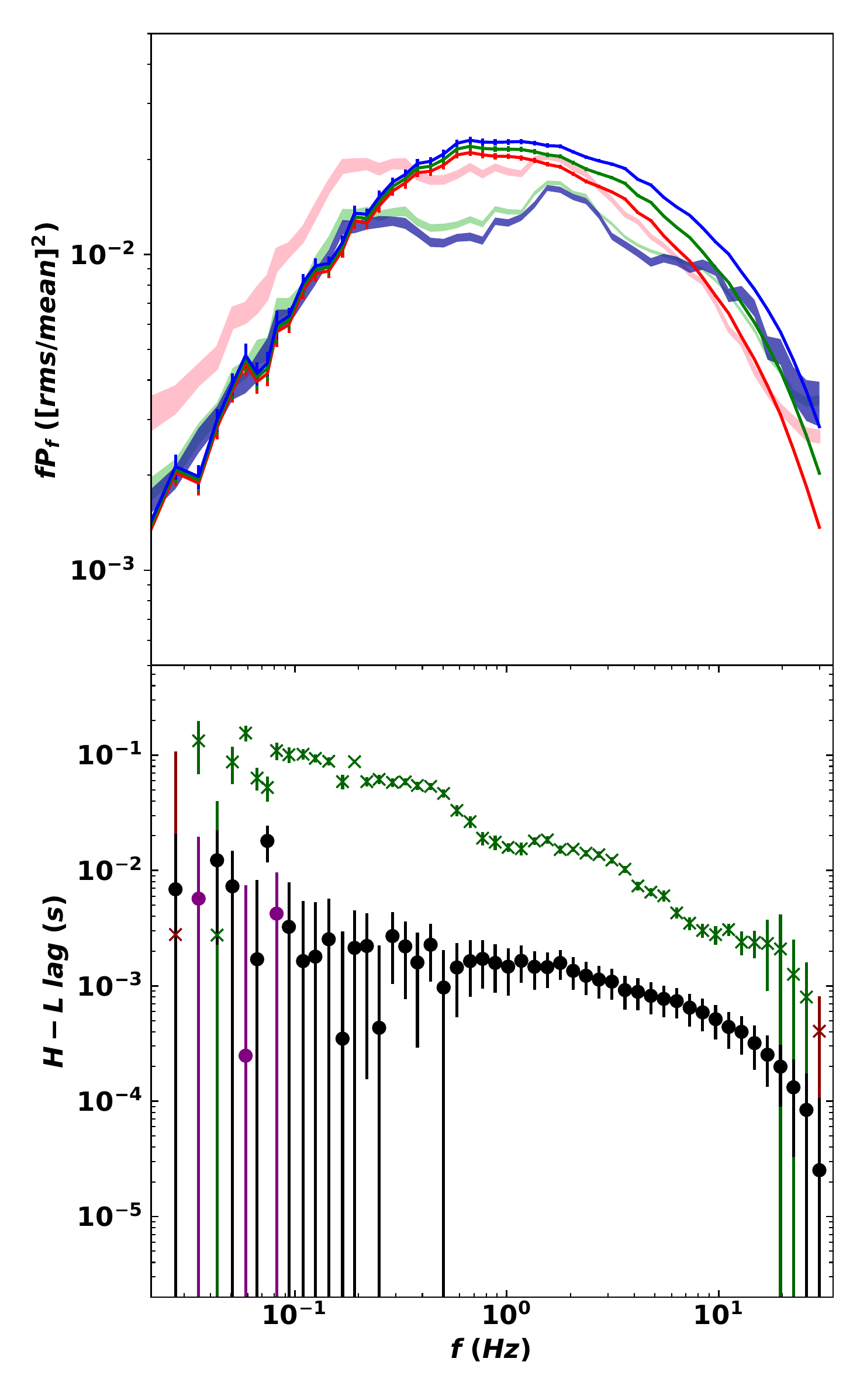}%
	\caption{Comparison to data for model where the timescale for fluctuations generation is $2\pi t_{visc}(r_n)$ instead of the usual $t_{visc}(r_n)$, while the propagation time remains $d\tau_n=t_{visc}(r_n)dr_n/r_n$. Top (a): High, Mid \& Low band PSDs. Colours as in Fig.~\ref{fig:pureprop_PSDs}. Bottom (b): High-Low band time lags. Colours as in Fig \ref{fig:pureprop_LAGS}.}
	\label{fig:sametimescaleproof2comp_2pi_ALLTHREE}
\end{figure}

So far, we have assumed that fluctuations are produced on the same timescale at which they propagate. We now decouple the generating timescale from the propagation timescale in order to determine the effect this has on the time lags. Section~\ref{The Propagating Fluctuations Model} argued that the largest-scale coherent fluctuations are generated over distances $h\sim r$, so that the maximum coherent timescale was $t_{visc}(r)$. However considering the flow over all azimuths means that a better estimate for the generating timescale would be the timescale of fluctuations with are coherent around the entire annulus, i.e. $2\pi t_{visc}(r_n)= 2\pi/f_{visc}(r_n)$, rather than $1/f_{visc}(r_n)$ as assumed thus far. Eq.~\ref{eq:Lorentzian} then becomes
\begin{equation}
\label{eq:TwoTimescaleLorentzian}
|\tilde{\dot{m}}(r_n, f)|^2 \propto \frac{1}{1+[2\pi f/f_{visc}(r_n)]^2}\left[\frac{sin(\pi f dt)}{\pi f dt}\right]^2,
\end{equation}
while the propagation timescale remains $d\tau_n=dr_n/r_nf_{visc}(r_n)$.

We assume the fiducial source size with $r_o=14$, $r_i=2.5$ and $r_{SH}=5.4$, and attempt to find a viscosity prescription which recovers a PSD similar to that in Fig.~\ref{fig:pureprop_PSDs}b (which has $B=250$ and $m=3.95$). As the generating frequency is now $f_{visc}(r_n)/2\pi$, instead of $f_{visc}(r_n)$, we start by dividing $B$ by $2\pi$ and leaving $m$ unchanged, but the altered effect of smoothing means that there is a better match for slightly different parameters, with $B=2\pi \times 51.83$ and $m=3.35$, and $F_{var}= 0.69$. Fig.~\ref{fig:sametimescaleproof2comp_2pi_ALLTHREE} shows the PSD and lags from this model. Interestingly, this has an even worse match to the observed lags than Fig.~ \ref{fig:pureprop_LAGS}b, because the lags are even shorter relative to the generating timescale than before.

An alternative approach one might consider would be to generate fluctuations on the  10$t_{kep}(r)$ timescale suggested by GRMHD simulations (\citealt{FM09}; \citealt{HR17}), while still propagating on $t_{visc}(r)$ set by the $f_{QPO}-f_b$ relation. However, generating on a 10$t_{kep}(r)$ timescale would require an inner flow radius much smaller than the ISCO size of ~2.5 $R_g$ in order to produce the required high-frequency power, so we do not explore this here.

In summary, the only case which approaches both the low-frequency PSD and low-frequency lags is the fiducial model in Figs.~\ref{fig:pureprop_PSDs}a and \ref{fig:pureprop_LAGS}a, whereby the generating and propagation timescales are set equal to $t_{visc}(r)$, derived from assuming that the QPO originates as Lense-Thirring precession of the hot flow. However, this model fails to explain the observed power at higher frequencies. This higher frequency power is also concentrated in a distinct `hump' around 2~Hz, unlike the smooth PSD produced in the propagation models with faster viscous timescales. Pure propagation models with self similar fluctuations cannot produce such humps, and neither can they explain how the Low-energy band can have more power than the Mid and High bands at most frequencies. We now explore a new family of models which allow the fractional variability to vary with radius, to see whether these can reproduce those essential features of the data.

\begin{figure}
	\includegraphics[width=\columnwidth]{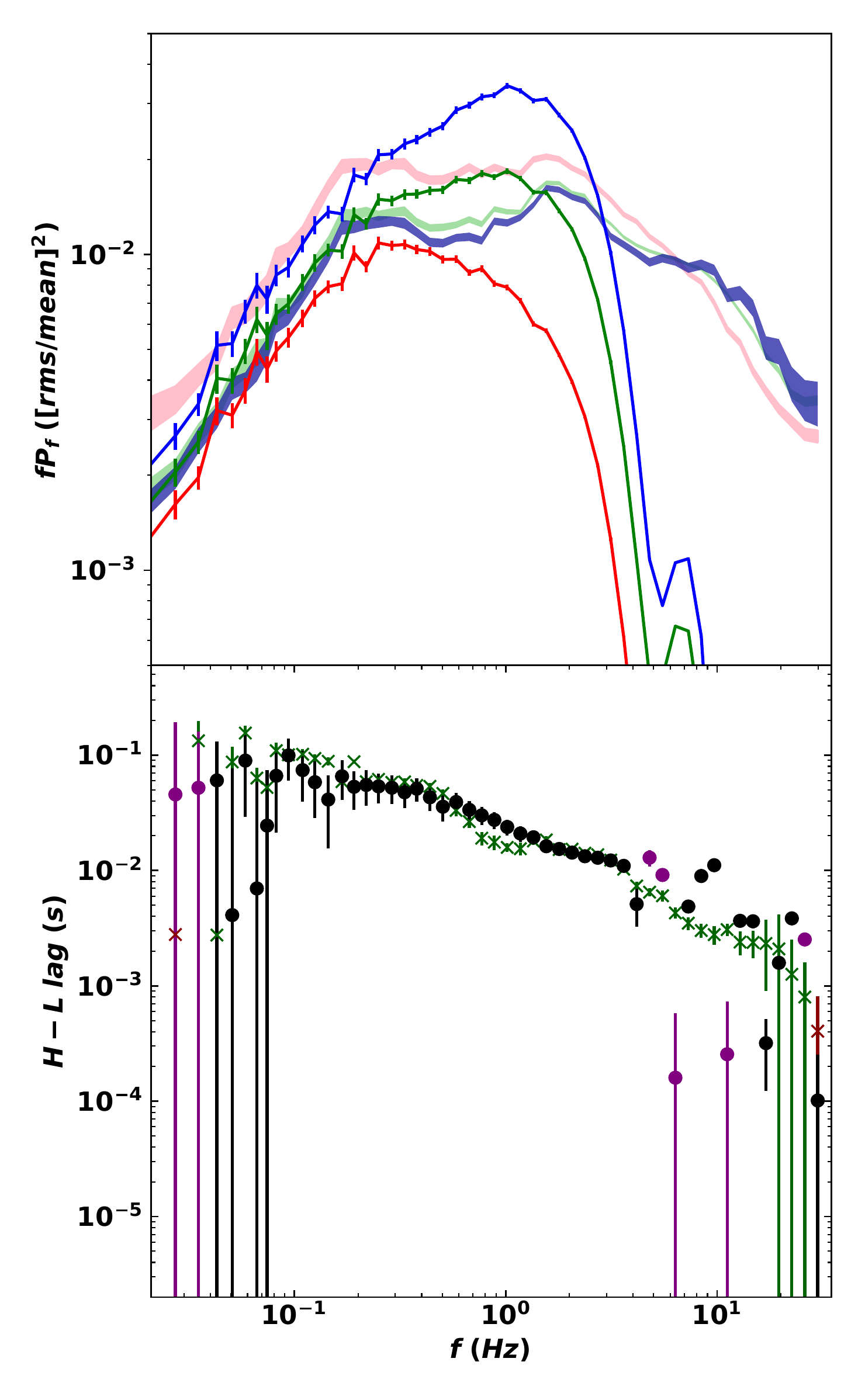}%
	\caption{The fiducial model ($B=0.03$, $m=0.5$, $r_o=14.$, $r_i=2.5$ $\gamma=3$ and the SF inner boundary), with additional variability at $r_{\alpha}$ such that $F_{var}(r_{\alpha})=14$. Top (a): High, Mid \& Low band PSDs. Colours as in Fig.~\ref{fig:pureprop_PSDs}. Bottom (b): High-Low band time lags. Colours as in Fig \ref{fig:pureprop_LAGS}.}
	\label{fig:ID11_stressfree_em3_1_radd_ALL2}
\end{figure}
\begin{figure}
	\includegraphics[width=\columnwidth]{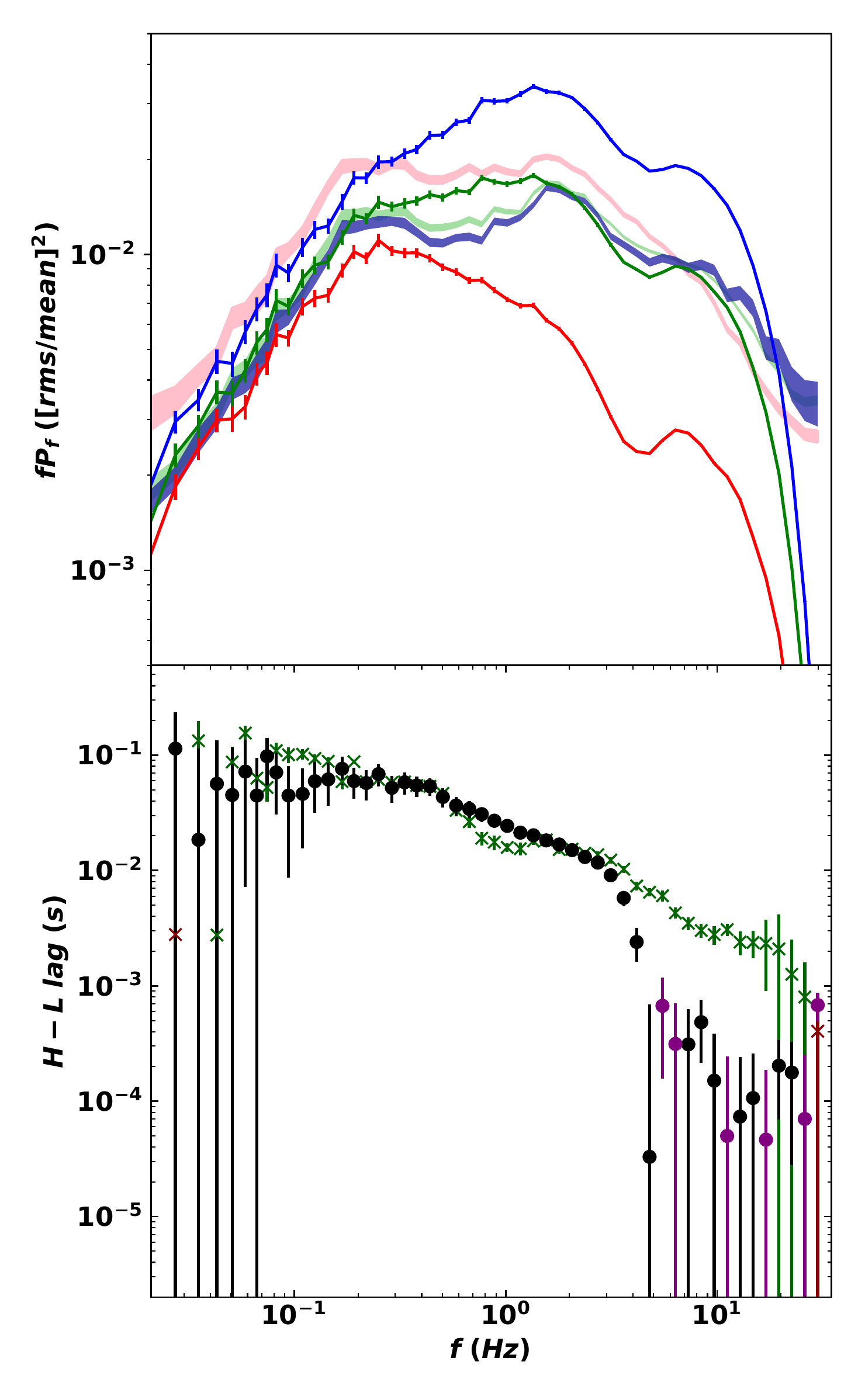}%
	\caption{The fiducial model ($B=0.03$, $m=0.5$, $r_o=14.$, $r_i=2.5$ $\gamma=3$ and the SF inner boundary), with additional variability at $r_{\alpha}$ and $r_{\beta}$, such that $F_{var}(r_{\alpha})$=$9$ and $F_{var}(r_{\alpha})$=$166$. Top (a): High, Mid \& Low band PSDs with colours as in Fig.~\ref{fig:pureprop_PSDs}. Bottom (b): High-Low band time lags with colours as in Fig \ref{fig:pureprop_LAGS}.}
	\label{fig:ID11_stressfree_em3_2_radds_ALL2}
\end{figure}
\section{Variability as a function of radius}

\label{AdVM}
The power spectra of the data are inherently `bumpy', and this is a generic feature in both Cyg X-1 (\citealt{CGR01}; \citealt{A08}; \citealt{T11}; G14) and other sources, such as GX 339-4 (\citealt{N00}). \citet{V16} proposes an idealised model to explain the bumpy PSDs from the interference of two radially separated, lagged Compton continua. However our results so far have shown that it becomes much more difficult for interference to produce the observed peaks if we consider the extended nature of the source and the generation of fluctuations at all radii. Alternatively, R16 suggest that the hump structure can be produced by considering fluctuations in the truncated thin disc at $r_o$. However the frequencies of these humps at $0.2$ or $2$~Hz are not easily consistent with any expected thin disc timescale. Instead the hump frequencies are more compatible with the viscous timescale within the flow itself. We therefore adapt our radially stratified model to allow enhanced variability at specific radii in the flow, in order to reproduce the observed PSD structure.

We keep the fiducial prescriptions for the viscous frequency ($B=0.03$, $m=0.5$, $r_o=14$, $r_i=2.5$), and emissivity ($\gamma=3$ and the SF boundary condition) as these match well to the observed low-frequency hump. $F_{var}$ is then allowed to vary with radius, so that the additional variability can be incorporated.

We first assume that there is enhanced turbulence at a specific radius $r_{\alpha}$, and derive this radius from the viscous frequency of the second peak in the PSD, i.e. $f_{visc}(r_{\alpha})=2$~Hz. From this we obtain $r_{\alpha}=5.5$, placing it  at the inner edge of the soft region since $r_{SH}=5.4$. All annuli in the flow apart from the one containing $r_{\alpha}$ have $F_{var}(r \neq r_{\alpha})=0.52$. The one which contains $r_{\alpha}$ requires $F_{var}(r_{\alpha})=14$ in order to match the amplitude of the 2~Hz peak in the Mid band PSD.

Fig.~\ref{fig:ID11_stressfree_em3_1_radd_ALL2}a shows that this additional power produces a divergence of the PSDs in different energy bands, reaching a maximum amplitude difference at $f_{visc}(r_{\alpha})=2$~Hz. However it is worth noting that since this is a log-log plot, the Mid and High bands are actually much less distinct than the Low. The divergence arises because $r_{\alpha}$ is situated close to $r_{SH}$, so only a small fraction of the soft region is affected by this additional variability, whereas it all propagates through the hard region. The Mid and High bands sample mostly from the hard region, while the Low band samples mostly from the soft region, resulting in this deficiency in power at high frequencies in the Low band.

Adding variability at $r_{\alpha}$ also does not reproduce the desired `hump' structure at 2~Hz. Instead the model PSDs are smooth from $f_b$ to $2$~Hz. This is due to propagation, since the noise generated in the soft region propagates coherently to $r_{\alpha}$, and so adds constructively to the additional noise. To obtain the observed decrease in the PSD from $0.3-1$~Hz requires that fluctuations are damped as they propagate, even more strongly than the smoothing in Eq.~\ref{smoothing} (see also R17).

The enhanced fluctuation power at $r_{\alpha}$ also underpredicts the high-frequency power above $2$~Hz, which rises to a third hump at $8$~Hz. This third Lorentzian peak is commonly seen in the power spectra of Cyg X-1 (\citealt{P03}; \citealt{A08}; Axelsson \& Done, in preparation), and potentially in other sources (e.g. GX 339-4; \citealt{N00}) indicating that the process driving this additional noise may be a fundamental physical mechanism in the Comptonising region.

We therefore add a second enhanced variability component at $r_\beta$ with amplitude $F_{var}(r_\beta)$ to match the third Lorenzian peak at 8~Hz. However, $r_\beta$ cannot now simply be derived from $f_{visc}(r_{\beta})=8$~Hz due to the increased effects of interference. Instead we fit for this, and find a best fit to the Mid-band PSD for these parameters of $F_{var}(r_{\beta})=166$ and $r_{\beta}=2.7$. The resulting PSDs are shown in Fig.~\ref{fig:ID11_stressfree_em3_2_radds_ALL2}. The Mid band power spectrum is now fairly well matched (apart from the dip between 0.3-1~Hz), as are the lags, but the Low- and High-energy PSDs are far from the observed data.

These results collectively support a model which relies on additional turbulence at characteristic positions in the flow to produce the high-frequency observed power. However certain key features have yet to be reproduced. The models so far have assumed that all variability from the outer regions is propagated, uninterrupted, into the inner regions such that Eq. \ref{pureprop} is applicable throughout the flow. However, the observed drop in power in the 0.3-1~Hz requires damping of the fluctuations. We now explore whether this can also finally reproduce the dominance of the Low-energy power spectrum over the High and Mid bands, whilst maintaining the lags of Fig.~\ref{fig:ID11_stressfree_em3_2_radds_ALL2}.

\begin{figure}
	\includegraphics[width=\columnwidth]{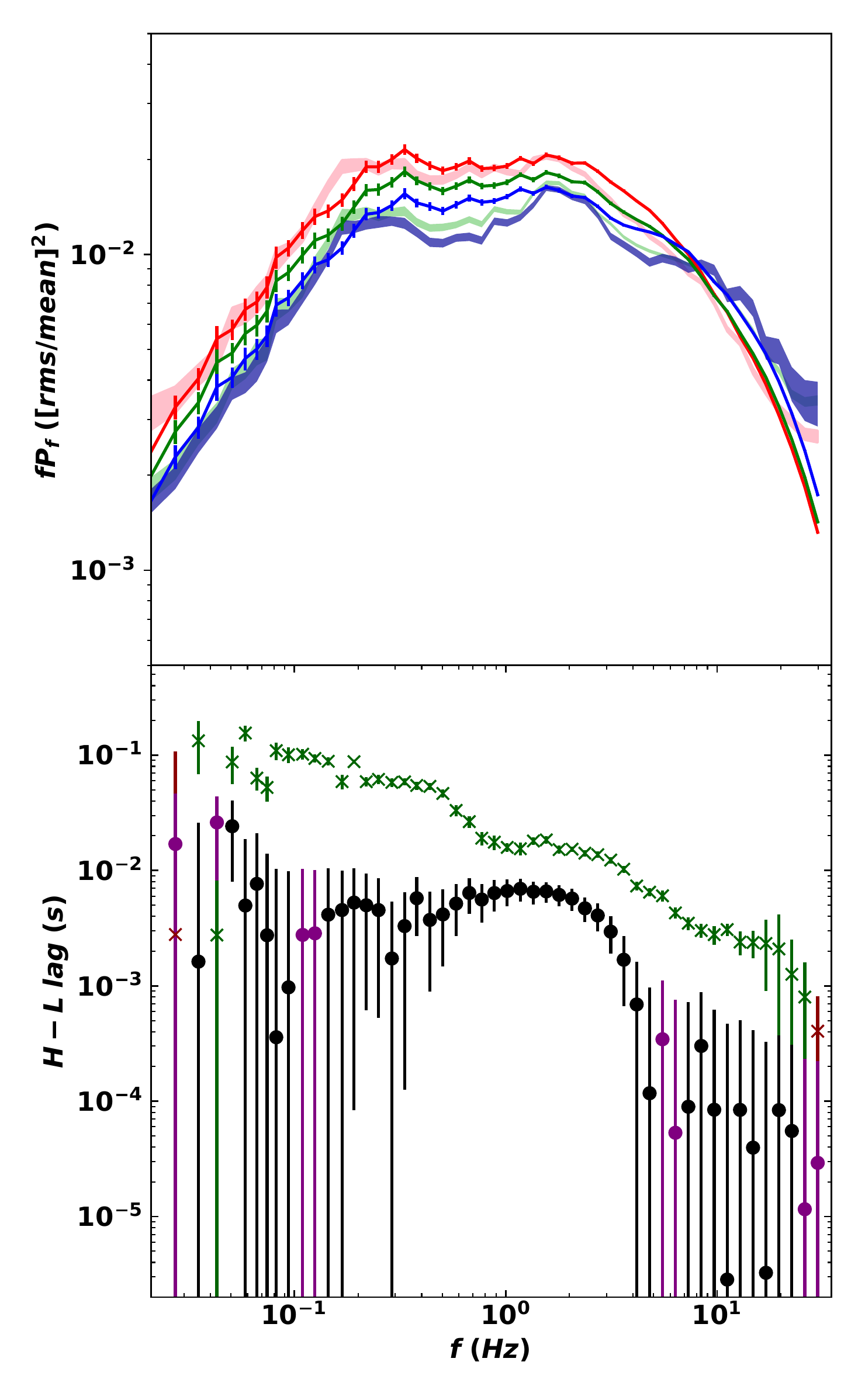}%
	\caption{Model as in Fig.~\ref{fig:ID11_stressfree_em3_2_radds_ALL2}, now with additional soft-component variability which is suppressed upon propagation. See text and Table \ref{tab:params} for parameters. Top (a): High, Mid \& Low band PSDs. Colours as in Fig.~\ref{fig:pureprop_PSDs}. Bottom (b): High-Low band time lags. Colours as in Fig \ref{fig:pureprop_LAGS}.}
	\label{fig:softsuppress_2comp_ALL2}
\end{figure}

\section{Damped Soft Variability}
\label{SoftPower}

The observed dominance of the Low-energy power has been seen in previous studies of Cyg X-1 (\citealt{G14}), as well as in other BHBs including SWIFT J1753.5-0127 and GX 339-4 (\citealt{WU09}), so it is a generic feature of the data.

Total propagation has so far prevented the Low-energy band from having more variability power at any frequency than the High band, since the low-frequency variability power dominating the Low band modulates the high-frequency power which dominates the High band. To suppress the transfer of low-frequency power to the High band then requires that some fraction of the variability power in the soft region fails to propagate into the hard region. However, we also require that the coherence between the fluctuations in each region is maintained. The soft variations must therefore map onto the inner-region variability after propagation, although with smaller amplitude and a time delay.

Physically, unpropagated noise could arise if part of the variability comes from disc seed-photon fluctuations. If the soft Comptonisation region becomes optically thick then it would shield the hard Comptonisation region from this variability component. Alternatively, part of this seed photon variability could be produced by a turbulent, clumpy transition between the truncated disc and hot flow, perhaps induced by instabilities in a shearing layer between the Keplerian disc and sub-Keplerian flow. These clumps might then evaporate, or be shredded by the MRI turbulence as they propagate inwards.

\begin{table*}
	\centering
	\begin{tabular}{lcccccccccccccr}
		\hline
		PSDs & Lags& $\gamma$ & $b(r)$ & $B$ & $m$ & $r_o$ & $r_i$ & $r_{SH}$& $r_{\alpha}$& $r_{\beta}$ & $F_{var}(r\neq r_{\alpha},  r_{\beta})$&$F_{var}(r_{\alpha})$& $F_{var}(r_{\beta})$ & $\cal{D}$ \\
		\hline
		Fig.~\ref{fig:ID11PSD} & - & 4.5 & 1 & 0.03 & 0.5 & 14. & 2.5 & 3.1 & - & - & 0.45& 1 & 1 & -  \\
		Fig.~\ref{fig:pureprop_PSDs}a & Fig.~\ref{fig:pureprop_LAGS}a & 3. & SF & 0.03 & 0.5 & 14. & 2.5 & 5.4 & - & - & 0.59& 1 & 1 & -  \\
		Fig.~\ref{fig:pureprop_PSDs}b & Fig.~\ref{fig:pureprop_LAGS}b & 3. & SF & 250.00 & 3.95 & 14. & 2.5 & 5.4 & - & - & 0.74& 1 & 1 & -  \\
		Fig.~\ref{fig:pureprop_PSDs}c & Fig.~\ref{fig:pureprop_LAGS}c & 3. & SF & 94.87 & 1.21 & 140. & 6. & 16.0 & - & - & 0.49& 1 & 1 & -  \\
		Fig.~\ref{fig:sametimescaleproof2comp_2pi_ALLTHREE}a & Fig.~\ref{fig:sametimescaleproof2comp_2pi_ALLTHREE}b & 3. & SF & $2\pi \times 51.83^*$ & 3.35 & 14. & 2.5 & 5.4 & - & - & 0.69& 1 & 1 & -  \\
		Fig.~\ref{fig:ID11_stressfree_em3_1_radd_ALL2}a& Fig.~\ref{fig:ID11_stressfree_em3_1_radd_ALL2}b & 3. & SF & 0.03 & 0.5 & 14. & 2.5 & 5.4 & 5.5 & - & 0.52& 14 & 1 & -  \\
		Fig.~\ref{fig:ID11_stressfree_em3_2_radds_ALL2}a& Fig.~\ref{fig:ID11_stressfree_em3_2_radds_ALL2}b& 3. & SF & 0.03 & 0.5 & 14. & 2.5 & 5.4 & 5.5 & 2.7 & 0.52& 9 & 166 & -  \\
		Fig.~\ref{fig:softsuppress_2comp_ALL2}a& Fig.~\ref{fig:softsuppress_2comp_ALL2}b& 3. & SF & 0.03 & 0.5 & 14. & 2.5 & 5.4 & 5.5 & 2.6 & $0.54^{**}$ & $35$ & $660$ & 60  \\
		\hline
	\end{tabular}
	\caption[caption]{Parameter values for all models shown in this work. \\\hspace{\textwidth}$^*$ This particular case decouples the fluctuation-generator and propagation timescales so that we still have $d\tau_n=dr_n/[r_nf_{visc}(r_n)]$ but now Eq. \ref{eq:TwoTimescaleLorentzian} describes the generator Lorentzians. \\\hspace{\textwidth}$^{**}$ Instead this is $F_{var}^S(r \neq r_{\alpha})$ as described in the text of $\S$ \ref{SoftPower}.}
	\label{tab:params}
\end{table*}

We model these effects generically by suppressing the amplitude of the propagated fluctuations by a factor, $\cal{D}$, at $r_{SH}$, and assume that the generated fluctuations within the hard region are also smaller by this factor than those generated in the soft region. For annuli without enhanced variability ($r \neq r_{\alpha}, r_{\beta})$, we therefore have $F_{var}^H=F_{var}^S/\cal{D}$.

\begin{figure}
	\includegraphics[width=\columnwidth]{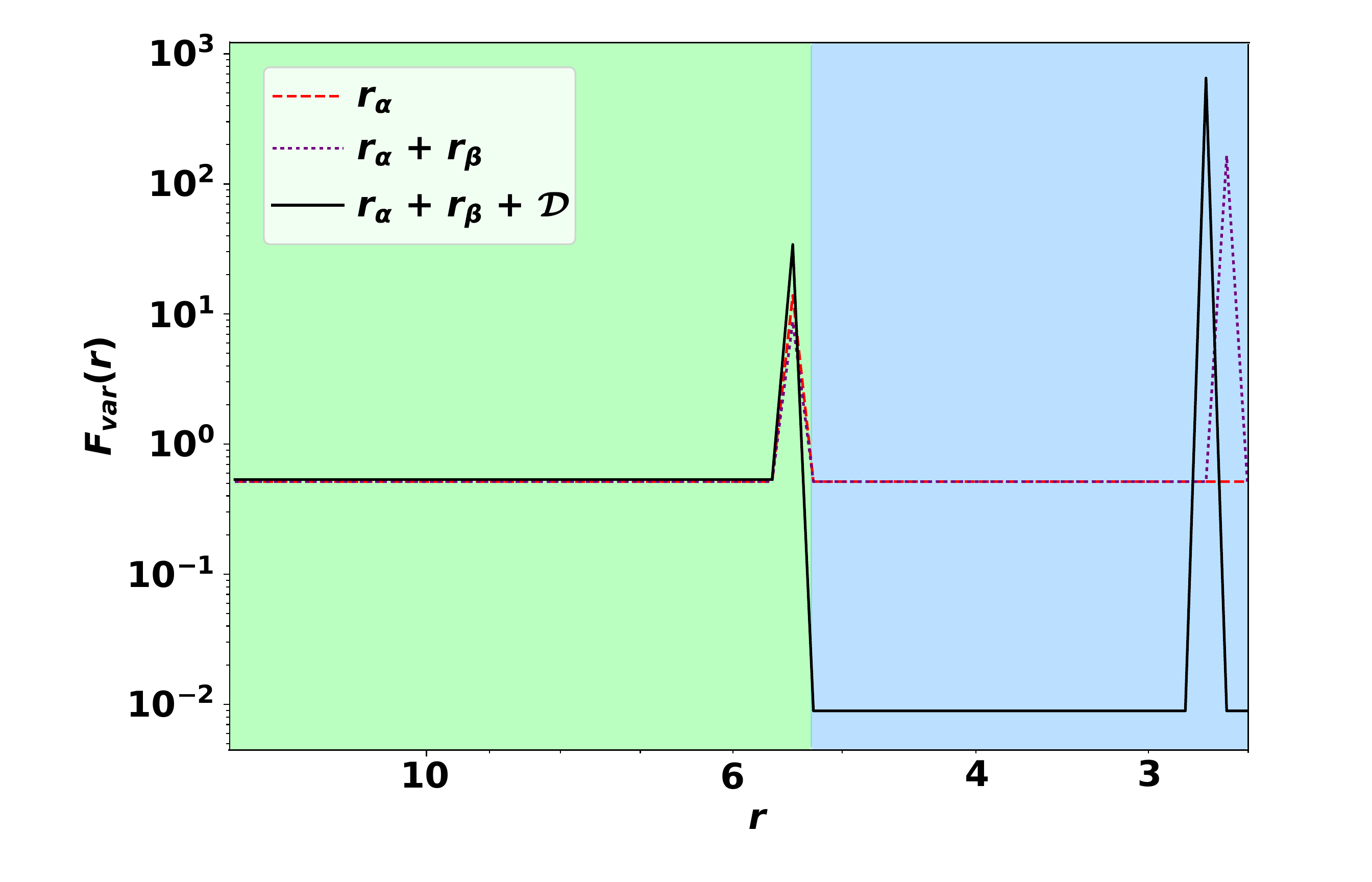}%
	\caption{The absolute fractional variability profiles used in the models for Figs.~\ref{fig:ID11_stressfree_em3_1_radd_ALL2}-\ref{fig:softsuppress_2comp_ALL2}. The green and turquoise shaded regions denote the soft and hard radial ranges of the flow respectively. The dashed red line denotes the $F_{var}$ profile of the model with enhanced variability only at $r_{\alpha}$ (Fig.~\ref{fig:ID11_stressfree_em3_1_radd_ALL2}). This radius is fixed such that $f_{visc}(r_{\alpha})=2$ Hz, at the second hump in the observed PSD, while  $F_{var}(r_{\alpha})$ here is a free parameter. The dotted purple line denotes that for the model with enhanced variability at $r_{\alpha}$ and $r_{\beta}$ (Fig.~\ref{fig:ID11_stressfree_em3_2_radds_ALL2}). For this case, $r_{\alpha}$, $r_{\beta}$ and their amplitudes, $F_{var}(r_{\alpha})$ and $F_{var}(r_{\beta})$, are allowed to be free due to the complications of interference. The solid black line is the $F_{var}(r)$ profile for the model of Fig.~\ref{fig:softsuppress_2comp_ALL2}, with $r_{\alpha}$, $r_{\beta}$, $F_{var}(r_{\alpha})$ and $F_{var}(r_{\beta})$ all set free. This model also includes damping of soft fluctuations propagating into the hard region and suppression of variability in the hard region, by a factor $\cal{D}$, also a free parameter.}
	\label{fig:Fvarprofiles}
\end{figure}

We use the fiducial frequency prescription ($B=0.03$, $m =0.5$) and size scale ($r_o=14$, $r_i=2.  5$) as this was the case which best approximated the observed lags. Fig.~\ref{fig:softsuppress_2comp_ALL2} shows the best fit found when the model is extended to include the free parameter, $\cal{D}$. An optimal fractional variability of $F_{var}^S=0.54$ is found on a simulation which also has additional variability at the two
characteristic radii of $r_{\alpha}=5.5$ and $r_{\beta}=2.6$ where $F_{var}^S (r_{\alpha})=35$ and $F_{var}^H (r_{\beta})=660$. The optimal suppression factor is found to be $\cal{D}=$~$60$. In Fig.~\ref{fig:Fvarprofiles}, we display the best fitting $F_{var}$ profiles of Sections~\ref{AdVM} and \ref{SoftPower} for ease of comparison. Using this parameterisation, we find the best approximation yet for the relative amplitudes and shapes of the PSDs in each energy band, although the difference in low-frequency power between the Low and Mid/High bands is still slightly underestimated.

However, Fig.~\ref{fig:softsuppress_2comp_ALL2}b shows that the simulated lags are now a poor match to the data, severely underestimating those which are observed, particularly at low frequencies. This is because the low frequency fluctuations are now highly damped, so they do not propagate sufficiently into the hard region for the cross-spectral lags to be significant. However, this prescription does reproduce the shape of the 2~Hz `step' in the lags, further suggesting the presence of specific radii in the flow which produce enhanced variability.

The other key shortfall of this model lies in the magnitudes of $F^S_{var}(r_{\alpha})$ and $F^H_{var}(r_{\beta})$. Large magnitude $F_{var}$ values such as these cause the generated light curves from these regions to go negative, which is clearly unphysical. Instead any future model demands smaller $F_{var}$ values in the regions of enhanced turbulence, so the emissivity in these regions must also be enhanced to transmit this smaller variability into the simulated light curves. This feature will be applied in future work (Mahmoud \& Done, in preparation). However the results we have shown here stand as a proof of concept that a non-uniform radial-variability profile is a key element in the timing behaviour of BHBs.

\section{Conclusions}
\label{conclusions}

We build a full spectral-timing model of fluctuations propagating down through a two component Comptonisation region in the BHB low/hard state. We systematically explore the effects of changing the model parameters on the energy dependent PSD and lags, and compare these to some of the best available data from Cyg X-1. We have fit to data only above 3~keV so that it is dominated by the flow, not by the intrinsic disc emission. The main results of this study can be summarised as follows:
\begin{enumerate}
\item The viscous frequency parameterisation is degenerate with the radial size scale of the Comptonising region. Time lags do not break this degeneracy without some external constraints from estimates of the truncated disc radius e.g. from spectral fitting of the broad iron line, a Lense-Thirring origin of the QPO, and/or light travel time lags. All of these support the ID11 prescription with $B=0.03$ and $m=0.5$, and so require that the low/hard state modelled here has an inner disc truncation radius of $\sim 14R_g$.
\item Coupling this to a standard emissivity with $\gamma=3$ and a stress-free inner boundary condition alone cannot produce the observed PSD using these parameters from a self-similar propagation model. This emissivity weights the observed power strongly to larger radii and hence lower frequencies, such that the significant variability observed above 0.5~Hz cannot be produced by this viscosity prescription alone.
\item Additional high-frequency power can only be produced in these models by assuming that there is enhanced turbulence within the flow, varying as a function of radius. This is also likely required to replicate the `steps' in the lag-frequency spectrum.
\item The PSD shape at all energies is emphatically non-monotonic, with a distinct dip in variability power between a low-frequency peak at 0.2~Hz and one at 2~Hz in these Cyg X-1 data. This distinct dip cannot be produced in any pure propagation model, and requires that variability from the outer flow is damped at some characteristic radius (or radii; see also R17).
\item The commonly observed low/hard state feature of Low-energy band dominance of the PSD at low frequencies requires that damping is included in the physical model. Some of the turbulence generated in the outer regions of the flow is not propagated down into the inner regions of the flow. However this damping also suppresses the lags.
\end{enumerate}

Our work adds to a growing understanding that the Comptonising region found in hard-state BHBs - far from being spectrally-homogeneous and smoothly variable - is almost certainly stratified in boths its spectrum and its variability (\citealt{WU09}; \citealt{V16}; R16; R17; \citealt{B17}). Clearly there are specific radii in the flow at which the fluctuations are enhanced and/or damped. These could be physically associated with the bending wave radius from a misaligned spinning black hole (\citealt{FM09}; \citealt{IDF09}), and/or the radius at which the jet is launched. A better understanding of the PSD and lags mean that we should be able to observationally trace the radii at which this physics operates. 

However, even with these additional model features, the energy-dependent PSDs and lags cannot be fit simultaneously with a two-Compton component spectral decomposition. Nonetheless, what we develop here is a flexible framework in which to construct a full spectral-timing model for the data. In future work we will modify this model to include more detailed spectral decompositions with three Compton components, which have been suggested by the most sophisticated spectral fits (\citealt{Y13}). We will also explore the effect of introducing a distinct fluctuation timescale at the disc-flow interface, to better approximate the variability in the potentially unstable disc-flow transition layer. 

\section*{Acknowledgements}
RDM acknowledges the support of an STFC studentship. RDM and CD thank Magnus Axelsson for helpful discussions on the spectral decompositions and double-peaked PSD. We also thank the anonymous referee for helpful comments which helped to improve the manuscript. This research has made use of data obtained through the High Energy Astrophysics Science Archive Research Center Online Service, provided by the NASA/Goddard Space Flight Center.

%%%%%%%%%%%%%%%%%%%%%%%%%%%%%%%%%%%%%%%%%%%%%%%%%%

%%%%%%%%%%%%%%%%%%%% REFERENCES %%%%%%%%%%%%%%%%%%

%%%%%%%%%%%%%%%%% APPENDICES %%%%%%%%%%%%%%%%%%%%%

\appendix
\section{Generalised Lags}
\label{AppendixCrossSpec}

IK13 show that the PSD form in Eq. \ref{eq:PSDprop2band} can be adapted to yield an analytic form for the cross spectrum between a low and high band, $\Gamma_{LH}(f)$. For $\dot{M}(r_n)$ with a mean of $\dot{M}_0$ and rms-normalisation, ignoring smoothing, this form becomes

\begin{equation}
\label{eq:}
\begin{aligned}
\Gamma_{LH}(f)=&\frac{1}{\mu_{L}\mu_{H}}\sum_{n=1}^{N}\left[w_n^{\,L}w_n^{\,H} P_{prop}(r_n,f) \right.\vphantom{}\\
 &+ \sum_{l=1}^{n-1}\left.\vphantom{} (w_l^{\,L}w_n^{\,H}e^{2\pi i \Delta\tau_{ln}f} +  w_l^{\,H} w_n^{\,L} e^{-2\pi i \Delta\tau_{ln}f})P_{prop}(r_l, f)\right],
\end{aligned}
\end{equation}
where
\begin{equation}
\mu_L=\sum_{r_n=r_i}^{r_o} \dot{M}_0 w_n^{L}  \text{ and } \mu_H=\sum_{r_n=r_i}^{r_o} \dot{M}_0 w_n^{H} .
\end{equation}
The real and imaginary parts respectively are then
\begin{equation}
\label{eq:CrossRe}
\begin{aligned}
\mathfrak{Re}[\Gamma_{LH}(f)]=&\frac{1}{\mu_{L}\mu_{H}}\sum_{n=1}^{N}\left[w_n^{\,L}w_n^{\,H}  P_{prop}(r_n,f)\right.\vphantom{}\\ &+ \sum_{l=1}^{n-1} \left.\vphantom{} \text{cos}(2\pi \Delta\tau_{ln}f)(w_l^{\,L}w_n^{\,H} + w_l^{\,H} w_n^{\,L})P_{prop}(r_l, f) \right],
\end{aligned}
\vspace*{-10pt}
\end{equation}
\begin{equation}
\label{eq:CrossIm}
\begin{aligned}
\mathfrak{Im}[\Gamma_{LH}(f)] =\frac{1}{\mu_{L}\mu_{H}}\sum_{n=1}^{N}\sum_{l=1}^{n-1}& \left[(w_l^{\,L}w_n^{\,H} - w_l^{\,H} w_n^{\,L})\right.\vphantom{}\\
&\left.\vphantom{}\times \text{sin}(2\pi \Delta\tau_{ln}f)P_{prop}(r_l, f)\right].
\end{aligned}
\end{equation}

\begin{figure}
	\includegraphics[width=\columnwidth]{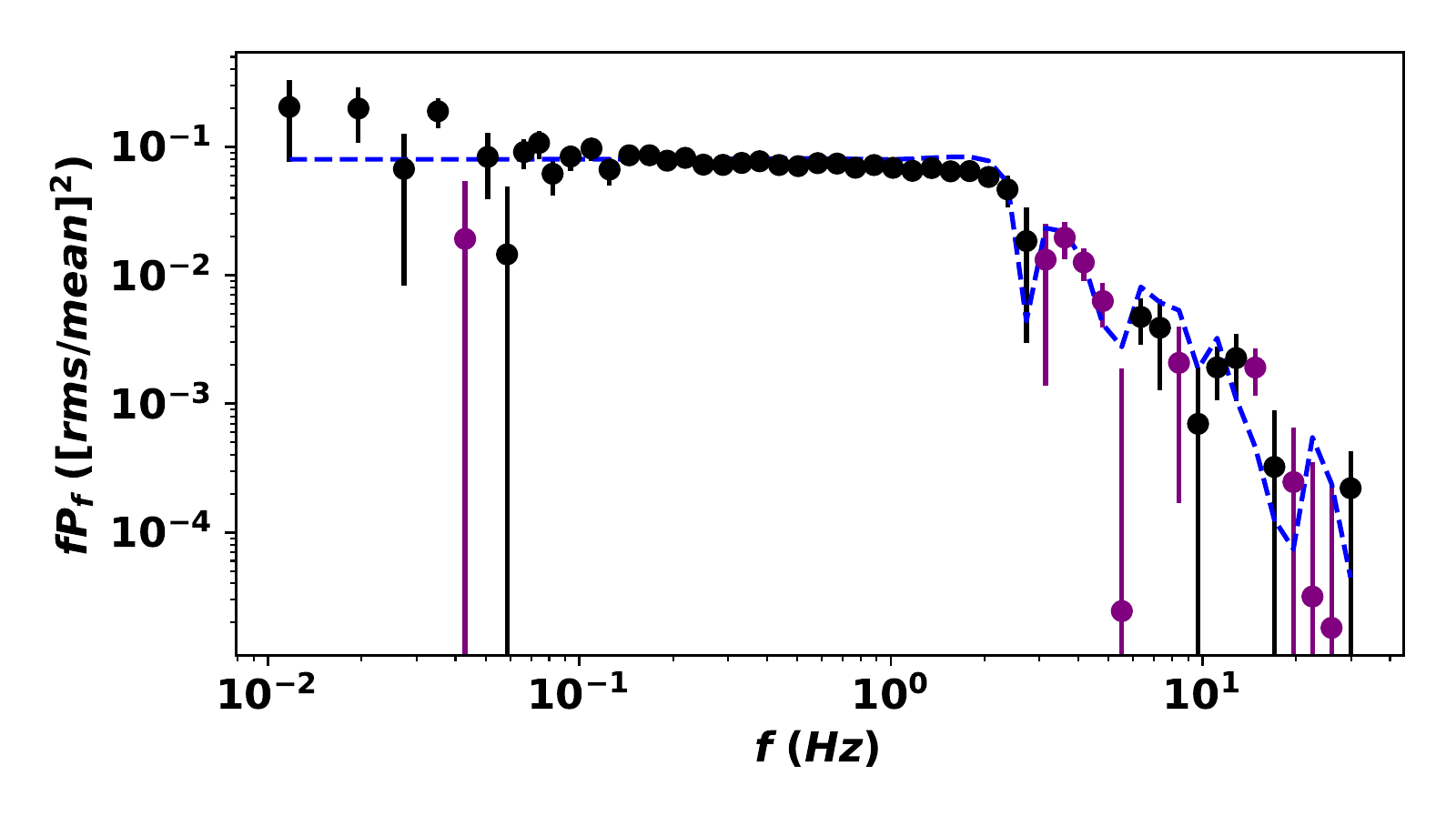}%
	\caption{Analytic (dashed blue line) comparison to simulated time lags for unsmoothed total propagation model with $B=0.03$, $m=0.5$, $r_o =14$, $r_i=2.5$, $\gamma =3$, $b(r)=3(1-\sqrt{r_i/r})$.}
	\label{fig:analag}
\end{figure}

From these components the time lag is computed in generality as
\begin{equation}
tan(2\pi f \tau_{meas})=\frac{\mathfrak{Im}[\Gamma_{LH}(f)]}{\mathfrak{Re}[\Gamma_{LH}(f)]}.
\end{equation}
We show an example of this analytic lag in Fig.~\ref{fig:analag}. Inconsistencies with the simulation output arise from the finite number of simulation realisations and the spatial resolution of the simulations.

\section{Analytic Case With Suppression of Soft Variability}

In the case of damping of the amplitude of the variability on propagation from the soft to the hard region, Eq. \ref{eq:PSDprop2band} modifies to 

\begin{equation}
\label{eq:PSDprop2bandmodded}
\begin{aligned}
P_{band}(f)=&\frac{1}{\mu_{C}^2}\sum_{n=1}^{N}\left[w_n^2 \frac{P_{prop}(r_n,f)}{d_n^2}\right.\vphantom{...}\\ 
&+ 2\sum_{l=1}^{n-1}\left.\vphantom{..} \frac{w_l w_n}{d_l d_n} \text{cos}(2 \pi \Delta\tau_{ln} f)P_{prop}(r_l, f)\right],
\end{aligned}
\end{equation}

where

\begin{equation}
d_m =
\begin{cases}
1 &\text{ if }\ r_m < r_{SH},\\
\cal{D} &\text{ if }\ r_m > r_{SH}.
\end{cases}
\end{equation}

The cross spectral components of Eqs. \ref{eq:CrossRe} and \ref{eq:CrossIm} also become

\begin{equation}
\label{eq:CrossReModded}
\begin{aligned}
\mathfrak{Re}[\Gamma_{LH}(f)]=&\frac{1}{\mu_{H}\mu_{L}}\sum_{n=1}^{N}\left[w_n^{\,H} w_n^{\,L} \frac{P_{prop}(r_n,f)}{d_n^2}\right.\vphantom{}\\ &+ \sum_{l=1}^{n-1} \left.\vphantom{} \text{cos}(2\pi \Delta\tau_{ln}f)(w_l^{\,L} w_n^{\,H} + w_l^{\,H}w_n^{\,L})\frac{P_{prop}(r_l, f)}{d_n d_l} \right],
\end{aligned}
\vspace*{-10pt}
\end{equation}
and
\begin{equation}
\label{eq:CrossImModded}
\begin{aligned}
\mathfrak{Im}[\Gamma_{LH}(f)] =\frac{1}{\mu_{H}\mu_{L}}\sum_{n=1}^{N}\sum_{l=1}^{n-1}& \left[(w_l^{\,L} w_n^{\,H} - w_l^{\,H}w_n^{\,L})\right.\vphantom{}\\
&\left.\vphantom{}\times \text{sin}(2\pi \Delta\tau_{ln}f)\frac{P_{prop}(r_l, f)}{d_n d_l}\right]
\end{aligned}
\end{equation}
respectively.

%%%%%%%%%%%%%%%%%%%%%%%%%%%%%%%%%%%%%%%%%%%%%%%%%%

\bsp
\label{lastpage}
\end{document}